\documentclass[%
 reprint,
superscriptaddress,
preprintnumbers,
 amsmath,amssymb,
 aps,
]{revtex4-2}

\usepackage{graphicx}
\usepackage{dcolumn}
\usepackage{bm}

\usepackage{aas_macros}
\usepackage{color}
\usepackage{amsmath}
\usepackage{natbib}

\usepackage{setspace}
\usepackage{subfigure}

\usepackage{hyperref}
\hypersetup{
    colorlinks=true,
    linkcolor=blue,
    citecolor=blue,        
    filecolor=magenta,      
    urlcolor=blue,
}

\usepackage{braket}

\begin{document}

\preprint{LA-UR-22-21601}

\title{
Non-iterative finite amplitude methods for E$1$ and M$1$ giant resonances
}


\author{Hirokazu Sasaki}
\email{hsasaki@lanl.gov}

\affiliation{Theoretical Division, Los Alamos National Laboratory, Los Alamos, New Mexico 87545, USA}
\author{Toshihiko Kawano}

\affiliation{Theoretical Division, Los Alamos National Laboratory, Los Alamos, New Mexico 87545, USA}
\author{Ionel Stetcu}

\affiliation{Theoretical Division, Los Alamos National Laboratory, Los Alamos, New Mexico 87545, USA}




\date{\today}

\begin{abstract}
The finite amplitude method (FAM) is a very efficient approach for solving the fully self-consistent random-phase approximation (RPA) equations. We use FAM to rederive the RPA matrices for general Skyrme-like functionals, calculate the electric dipole (E1) and the magnetic dipole (M1) giant resonances, and compare the results with available experimental and evaluated data. For the E1 transitions in heavy nuclei, the calculations reproduce well the resonance energy of the photoabsorption cross sections. In the case of M1 transitions, we show that the residual interaction does not affect the transition strength of double-magic nuclei, which suggests that the spin terms in the Skyrme force currently neglected in the present computation could improve the agreement between FAM and experimental data.

\end{abstract}

\maketitle


\section{Introduction}

Giant resonances are collective motions of many neutrons and protons in the atomic nucleus induced by weak external perturbations \cite{harakeh2001giant}. The electric dipole resonance (GDR) was first observed by Bothe and Gentner in their photo-absorption experiments \cite{Bothe:1937ZP}. Since then, various types of giant resonances are found in nuclear experiments \cite{Berman:1975tt,Speth:1981gdr,harakeh2001giant}. Because of the time-reversal invariance in the compound nucleus reaction, strengths of the electric dipole (E$1$) and the magnetic dipole (M$1$) excitations are translated into the photon emission process, namely the neutron capture cross sections \cite{Goriely:2003rz,Mumpower:2017gqj}, where an implicit Brink-Axel hypothesis~\cite{Brink:1955phd} is always assumed. The obtained capture cross sections can be applied to nucleosynthesis calculations in explosive astrophysical sites. It is theoretically predicted that large numbers of unstable nuclei are produced through nucleosynthesis of heavy elements such as the $r$-process \cite{Mumpower:2015ova,Kajino:2019abv,Cowan:2019pkx} and the $\nu p$-process \cite{Frohlich:2005ys,Sasaki:2017jry,Sasaki:2021ffa}. The unstable nuclei produced in these processes decay to stable nuclei, and they contribute to a significant fraction of heavy elements in the solar abundance. In induced or spontaneous fission, neutron-rich nuclei far from stability follow the same decay toward stability. Because the experimental data on giant resonances are mainly limited to stable nuclei, the reaction rates on unstable nuclei in nuclear network calculations or fission simulations inevitably depend on the theoretical prediction. Therefore, reliable theoretical calculations of giant resonances are essential to study the origin of heavy elements and not only. In addition, phenomenological models of GDR, which are often employed in calculating the reaction rates for neutron-rich targets, should be validated against more fundamental theoretical models.

As a contemporary theoretical approach, the nuclear density functional theory (DFT) is used to study the properties of the ground and excited states in the nuclear many-body system. Random-phase approximation (RPA) derived from the time-dependent Hartree-Fock (TDHF) equation can describe dynamical properties of nuclei for external fields \cite{ring2004nuclear}. RPA is extended to quasiparticle random-phase approximation (QRPA) by including the pairing correlations \cite{rowe2010nuclear}. Collective excitations like GDR have been extensively studied in the (Q)RPA calculations \cite{harakeh2001giant,Nakatsukasa:2016nyc,Colo:2022a}. Growing interest includes the M1 transitions, which has been studied in various (Q)RPA approaches \cite{Vesely:2009eb,Nesterenko:2010ra,Cao:2009hh,Goriely:2016flw,Tselyaev:2019ovd,Kruzic:2020oqf,Oishi:2019ilc,Nesterenko:2021wkm}. Because (Q)RPA is the small amplitude limit of time-dependent DFT, time-dependent Hartree-Fock-Bogoliubov (TDHFB) \cite{Ebata:2010qr} and time-dependent superfluid local density approximation (TDSLDA) \cite{Stetcu:2011kx} have been also used to calculate giant resonances.

Although DFT-based microscopic approaches have few adjustable parameters, which are usually fitted to masses and select other static properties, solving numerically the QRPA equations is computationally expensive, especially in the case of deformed superfluid nuclei. The finite amplitude method (FAM) is a much less demanding alternative \cite{Nakatsukasa:2007qj,Inakura:2009vs,Avogadro:2011gd,Stoitsov:2011zz,Stoitsov:2011zz,Hinohara:2013qda,Kortelainen:2015gxa,Oishi:2015lph,Liang:2013pda,Niksic:2013ega,Avogadro:2013uf,Mustonen:2014bya,Shafer:2016etk,Ney:2020mnx,Hinohara:2022arXiv220112983H,Washiyama:2020qfr} to numerically solve the full self-consistent (Q)RPA equations. In FAM calculations, the linear RPA equations are solved iteratively. Because FAM avoids the diagonalization of the large size of (Q)RPA matrices, it reduces the computational costs significantly. FAM was first proposed within the RPA framework \cite{Nakatsukasa:2007qj} and was used to study the E1 transitions \cite{Inakura:2009vs}. FAM-RPA was soon extended to the QRPA calculations \cite{Avogadro:2011gd} and has been applied to various multipole modes in deformed nuclei \cite{Stoitsov:2011zz,Hinohara:2013qda,Kortelainen:2015gxa,Oishi:2015lph}. There are various applications of the FAM approach, e.g., the extension to the relativistic framework \cite{Liang:2013pda,Niksic:2013ega} and the construction of RPA matrices in the matrix FAM (m-FAM) \cite{Avogadro:2013uf}. Furthermore, the Skyrme proton-neutron FAM (pnFAM) was applied to the beta-decay \cite{Mustonen:2014bya,Shafer:2016etk,Ney:2020mnx} and the two-neutrino double-beta decay \cite{Hinohara:2022arXiv220112983H}. Finally,
FAM-QRPA was also efficiently used to calculate the collective inertia in fission dynamics \cite{Washiyama:2020qfr}.

In this paper, we analytically derive the contribution of residual interaction in the RPA equations and calculate cross sections of E$1$ and the M$1$ transitions. In our implementation of FAM, the residual interaction is explicitly linearized. We chose to solve the RPA matrix equations rather than following an iterative approach, as this involves simply the inversion of a matrix that can be performed numerically very efficiently. 

\section{Theory}

\label{sec:theory}

\subsection{Finite amplitude method (FAM)}

We briefly review the formalism of the FAM-RPA calculation following details in Ref.~\cite{Nakatsukasa:2007qj,Inakura:2009vs}. In the static Hartree-Fock (HF) calculation, we can obtain iteratively the static HF Hamiltonian and the associated single-particle states, $h_{0}\ket{\phi_{\mu}}=\epsilon_{\mu}\ket{\phi_{\mu}}$. The single-particle states are divided in occupied (hole) states $\left\{\phi_{i}\right\}(i=1,..., A)$ and unoccupied (particle) states $\left\{\phi_{m}\right\}(m=A+1,...)$. Hereafter, we use indices $i,j$ for occupied states and $m,n$ for unoccupied states. 
Nuclear excitations caused by a weak external field can be studied through the time evolution of the one-body density matrix $\rho(t)$ \cite{ring2004nuclear}. A time-dependent external field $V_{\rm ext}(t)$ causes the transition density $\delta\rho(t)$ and the associated residual field $\delta h(t)$. In the frequency space, these are denoted by $V_{\mathrm{ext}}(\omega)$, $\delta\rho(\omega)$, and $\delta h(\omega)$.
Assuming a weak external field, the transition density in the $\omega$ representation is expressed as
\begin{equation}
    \delta\rho(\omega) = \sum_{i=1}^{A}\left\{
    \ket{X_{i}(\omega)}\bra{\phi_{i}}+\ket{\phi_{i}}\bra{Y_{i}(\omega)}
    \right\},
\end{equation}
where $\ket{X_{i}(\omega)}$ ($\ket{Y_{i}(\omega)}$) are the forward (backward) amplitudes. From the linear response of the TDHF equations, the linear RPA equations are given by
\begin{equation}
\label{eq:RPA linear response X}
\begin{split}
\omega\ket{X_{i}(\omega)} &= \left(
h_{0}-\epsilon_{i}
\right)\ket{X_{i}(\omega)} \\
&+ \hat{P}
\left\{
V_{\mathrm{ext}}(\omega) + \delta h(\omega)
\right\}\ket{\phi_{i}},
\end{split}
\end{equation}
\begin{equation}
\label{eq:RPA linear response Y}
\begin{split}
-\omega\bra{Y_{i}(\omega)} &= \bra{Y_{i}(\omega)}\left(
h_{0}-\epsilon_{i}
\right)\\
&+ \bra{\phi_{i}}\left\{
V_{\mathrm{ext}}(\omega) + \delta h(\omega)
\right\}\hat{P},
\end{split}
\end{equation}
where $\hat{P}=1-\sum_{i=1}^{A}\ket{\phi_{i}}\bra{\phi_{i}}$ is the projector onto unoccupied states. In the general framework of FAM, the residual interaction is linearized by a small parameter $\eta$ and written as
\begin{equation}
\label{eq:FAM}
    \delta h(\omega)=\frac{h(\bra{\psi^{\prime}_{i}},\ket{\psi_{i}})-h_{0}}{\eta},
\end{equation}
where $h$ is the single-particle Hamiltonian of the TDHF calculation and $\ket{\psi_{i}}=\ket{\phi_{i}}+\eta\ket{X_{i}(\omega)}$,   $\bra{\psi_{i}^{\prime}}=\bra{\phi_{i}}+\eta\bra{Y_{i}(\omega)}$ are the ket and bra representations of the single-particle state of $h$. The residual interaction in Eq.~(\ref{eq:FAM}) is obtained numerically by employing the small parameter $\eta$ and the single-particle state. Equations (\ref{eq:RPA linear response X}) and (\ref{eq:RPA linear response Y}) are solved iteratively together with Eq.~(\ref{eq:FAM}).

 \indent The FAM calculation is often carried out in the mixed representation \cite{Inakura:2009vs,Oishi:2015lph}. However, in this paper, we solve the linear RPA equations in the matrix form. The forward and backward amplitudes are decomposed by the unoccupied states \cite{Nakatsukasa:2007qj},
\begin{equation}
\label{eq:X in matrix form}
\ket{X_{i}(\omega)} = \sum_{m>A}X_{mi}(\omega)\ket{\phi_{m}},
\end{equation}
\begin{equation}
\label{eq:Y in matrix form}
\ket{Y_{i}(\omega)} = \sum_{m>A}Y_{mi}^{*}(\omega)\ket{\phi_{m}}.
\end{equation}
In the matrix form, the RPA equations in Eqs.~(\ref{eq:RPA linear response X}) and (\ref{eq:RPA linear response Y}) are described by
\begin{eqnarray}
  (\epsilon_{m}-\epsilon_{i}-\omega) X_{mi}(\omega)
      &+& \bra{\phi_{m}}\delta h(\omega)\ket{\phi_{i}} \nonumber\\
  &=& -\bra{\phi_{m}}V_{\mathrm{ext}}(\omega)\ket{\phi_{i}} \ ,
  \label{eq:RPA linear response X in matrix form} \\
  (\epsilon_{m}-\epsilon_{i}+\omega) Y_{mi}(\omega)
      &+& \bra{\phi_{i}}\delta h(\omega)\ket{\phi_{m}} \nonumber \\
  &=& -\bra{\phi_{i}}V_{\mathrm{ext}}(\omega)\ket{\phi_{m}} \ .
  \label{eq:RPA linear response Y in matrix form}
\end{eqnarray}
In order to simplify the notation, the index $\omega$ is dropped hereafter. The external field and the residual interactions are independent of $\omega$ in our numerical calculation. In the following discussion, it is assumed that the external field $V_{\mathrm{ext}}$ does not change the isospin of the nucleon.

\subsection{RPA equations}
\label{sec:TDHF Hamiltonian and the RPA matrices}

The RPA matrices $A$ and $B$ in the well-known RPA equation \cite{ring2004nuclear} are derived from the explicit linearization of the TDHF Hamiltonian with the expansion parameter $\eta$. Because we work with Skyrme-type nuclear energy density functionals the single-particle Hamiltonian writes \cite{Engel:1975zz,Dobaczewski:1995zz,Maruhn:2013mpa,Jin:2020kdn}
\begin{equation}
\begin{split}
h=\sum_{q}h_{q}=\sum_{q}\left(
h_{q}^{\mathrm{even}}+h_{q}^{\mathrm{odd}}
\right),
\end{split}
\end{equation}
\begin{eqnarray}
\label{eq:TDHF Hamiltonian even}
  h_{q}^{\mathrm{even}}
  &=& -\vec{\nabla}\cdot\frac{\hbar^{2}}{2m^{*}_{q}}\vec{\nabla} + U_{q}+\vec{W}_{q}\cdot\left(
    -i
    \right)\left(
    \vec{\nabla}\times\vec{\sigma}
    \right) \nonumber \\
    &+& \delta_{qp}V_{\mathrm{coul}},
\end{eqnarray}
\begin{equation}
\label{eq:TDHF Hamiltonian odd}
h_{q}^{\mathrm{odd}}=\vec{S}_{q}\cdot\vec{\sigma}-\frac{i}{2}\left[
(\vec{\nabla}\cdot \vec{A}_{q})+2\vec{A}_{q}\cdot\vec{\nabla}
\right],
\end{equation}
\begin{equation}
V_{\mathrm{coul}}=\frac{e^{2}}{2}\int\mathrm{d}^{3}r^{\prime}\ \frac{\rho_{p}}{|\vec{r}-\vec{r^{\prime}}|}-\frac{e^{2}}{2}\left(
\frac{3}{\pi}
\right)^{1/3}\rho_{p}^{1/3},
\end{equation}
where $\vec{r}$ is the space coordinate and $\vec{\sigma}$ represents the spin of the nucleon $q$  ($q=n$ for neutrons and $q=p$ for protons), while $h_{q}^{\mathrm{even}}$ and $h_{q}^{\mathrm{odd}}$ are the time-even and time-odd Hamiltonian, respectively. The central potential $U_{q}$, the effective mass $m^{*}_{q}$, and the spin-orbit potential $\vec{W}_{q}$ in the time-even Hamiltonian are calculated using the nucleon particle density $\rho_{q}$, the kinetic energy density $\tau_{q}$, and the spin-orbit density $\vec{J}_{q}$ \cite{Engel:1975zz,Maruhn:2013mpa}. These local quantities are computed with the single-particle states of the TDHF Hamiltonian $\{\psi_{\mu},\psi_{\mu}^{\prime *}\}$ (see the detail in Sec.\ref{sec:residula time-even}). $V_{\mathrm{coul}}$ is the Coulomb potential, where $\rho_{p}$ is the proton density defined in Eq.~(\ref{eq:time-even fields}). For even-even nuclei, the time-odd Hamiltonian does not contribute to the static HF calculation because the time-odd fields such as the spin density $\vec{s}_{q}$ and the current density $\vec{j}_{q}$ used to compute $\vec{S}_{q}$ and $\vec{A}_{q}$ are zero under the time-reversal symmetry \cite{Engel:1975zz,Maruhn:2013mpa}. Hence, the single particle states $\{\phi_{\mu}\}$ are determined solving the eigenvalue equation for $h_{q}^{\mathrm{even}}$. On the other hand, the density matrix cannot be time-even in the dynamical evolution. The time-odd contribution must be included for even for even-even nuclei in order to satisfy the Galilean invariance \cite{Engel:1975zz,Jin:2020kdn}. In the  FAM-RPA calculation, both time-even and time-odd potentials are expressed as the functions of the self-consistent single particle states without any mixing. According to Eqs.~(\ref{eq:X in matrix form}), (\ref{eq:Y in matrix form}), the occupied state and their complex conjugate of the TDHF Hamiltonian in our FAM calculation are expanded by the small parameter $\eta$,
\begin{eqnarray}
\label{eq:single-particle states TDHF}
\psi_{i}(\vec{r},\sigma,q)
  &=& \phi_{i}(\vec{r},\sigma,q) \nonumber \\
  &+& \eta\sum_{m\in q}X_{mi}^{q}\phi_{m}(\vec{r},\sigma,q)+\mathcal{O}(\eta^{2}) \ ,
\end{eqnarray}
\begin{eqnarray}
\label{eq:single-particle states TDHF complex conjugate}
\psi_{i}^{\prime*}(\vec{r},\sigma,q)
  &=& \phi_{i}^{*}(\vec{r},\sigma,q) \nonumber \\ 
  &+& \eta\sum_{m\in q}Y_{mi}^{q}\phi_{m}^{*}(\vec{r},\sigma,q)+\mathcal{O}(\eta^{2}) \ .
\end{eqnarray}
For a detailed discussion on how we calculate $\phi_{i}(\vec{r},\sigma,q)$ in our numerical implementation, we refer the interested reader to Appendix \ref{sec:detail of integrands}. In the limit of $\eta\to0$, the residual interaction in Eq.~(\ref{eq:FAM}) should be independent of $\eta$ and expressed as linear combinations of coefficients $\{X_{nj}^{q^{\prime}}\}$ and $\{Y_{nj}^{q^{\prime}}\}$,
\begin{eqnarray}
\label{eq:linearization of the residual interaction}
\lim_{\eta \to 0}\delta h 
  &=& \sum_{q^{\prime}}\sum_{nj\in q^{\prime}} 
        X_{nj}^{q^{\prime}}
        \left. \frac{\partial h}{\partial (\eta X_{nj}^{q^{\prime}})} \right|_{\eta=0} \nonumber \\
  &+& \sum_{q^{\prime}}\sum_{nj\in q^{\prime}}
        Y_{nj}^{q^{\prime}}
        \left. \frac{\partial h}{\partial (\eta Y_{nj}^{q^{\prime}})} \right|_{\eta=0}\ .
\end{eqnarray}
In such explicit linearization of the residual interactions, we no longer need the small parameter $\eta$ and an iterative procedure to solve Eqs.~(\ref{eq:RPA linear response X in matrix form}) and (\ref{eq:RPA linear response Y in matrix form}). The expansions of the single-particle states in Eqs.~(\ref{eq:single-particle states TDHF}) and (\ref{eq:single-particle states TDHF complex conjugate}) enable the explicit linearization of the Skyrme potentials. When the external field and the single-particle Hamiltonian are local in the coordinate space, the RPA equations in Eqs.~(\ref{eq:RPA linear response X in matrix form}) and (\ref{eq:RPA linear response Y in matrix form}) are described in the matrix form,
\begin{equation}
\left\{
\left(
\begin{array}{cc}
     A & B \\
     B^{*} & A^{*} 
\end{array}
\right)
-\omega
\left(
\begin{array}{cc}
     1 & 0 \\
     0 & -1
\end{array}
\right)
\right\}
\left(
\begin{array}{c}
     X_{nj}^{q^{\prime}}\\
     Y_{nj}^{q^{\prime}}
\end{array}
\right)
=-
\left(
\begin{array}{c}
     f_{mi}^{q}\\
     f_{im}^{q}
\end{array}
\right),
\label{eq:RPAeq}
\end{equation}
\begin{eqnarray}
A^{q,q^{\prime}}_{mi,nj}
  &=& \left(
        \epsilon_{m}-\epsilon_{i}
      \right) \delta_{mn}\delta_{ij} \nonumber \\
  &+& \int\mathrm{d}^{3}r\ \phi_{m}^{q*}
      \left(
        \frac{\partial h_{q}}{\partial (\eta X_{nj}^{q^{\prime}})} 
      \right)_{\eta=0}\phi_{i}^{q} \ , \label{eq:RPA matrix A}\\
B_{mi,nj}^{q,q^{\prime}}
  &=& \int\mathrm{d}^{3}r\ \phi_{m}^{q*}
      \left(
        \frac{\partial h_{q}}{\partial (\eta Y_{nj}^{q^{\prime}})}
      \right)_{\eta=0}\phi_{i}^{q} \ ,
\label{eq:RPA matrix B}
\end{eqnarray}
\begin{equation}
\label{eq:external field f}
f_{mi}^{q}=\int\mathrm{d}^{3}r\ \phi_{m}^{q*}V_{\mathrm{ext}}\phi_{i}^{q},\ \ f_{im}^{q}=\int\mathrm{d}^{3}r\ \phi_{i}^{q*}V_{\mathrm{ext}}\phi_{m}^{q},
\end{equation}
where the limit in Eq.~(\ref{eq:linearization of the residual interaction}) is used to configure the elements of RPA matrices in Eqs.~(\ref{eq:RPA matrix A}) and (\ref{eq:RPA matrix B}). In the equations above, we use a simple notation, $\phi^{q}_{\mu}\equiv\phi_{\mu}(\vec{r},\sigma,q)$. The matrices $A_{mi,nj}^{q,q^{\prime}}$ and $B_{mi,nj}^{q,q^{\prime}}$ are functions of $\phi_{i}^{q}$, $\phi_{m}^{q*}$, $\phi_{j}^{q^{\prime}*}$, and $\phi_{n}^{q^{\prime}}$ with the various Skyrme parameters \cite{Chabanat:1997un}. A detailed description of the RPA matrices $A$ and $B$ is given below.

The argument for solving iteratively the FAM equations is the size of the matrices $A$ and $B$, which is significant in the case of QRPA, especially for deformed nuclei. While this is correct, one can also argue that  extremely efficient numerical methods are available for inverting large matrices, especially if one uses a parallel algorithm. Thus, while we chose to use complex $\omega$, like in the other FAM-based approaches, we calculate explicitly the $A$ and $B$ matrices and by direct inversion, we solve for the amplitudes $X$ and $Y$ in Eq. (\ref{eq:RPAeq}). Since $A$ and $B$ are given at a one-time cost, the approach will most likely compete with iterative FAM.

\subsection{The residual interaction of $h_{q}^{\mathrm{even}}$}
\label{sec:residula time-even}
The RPA matrices $A$ and $B$ are composed of local densities and currents given by the single-particle states of the HF calculation. In our RPA calculation, the effective mass $m^{*}_{q}$ and potentials such as $U_{q}$ and $W_{q}$ in the time-even Hamiltonian are obtained from the nucleon density $\rho_{q}$, the kinetic energy density $\tau_{q}$, and the spin-orbit density $\vec{J}_{q}$ defined as \cite{Vautherin:1973zz},
\begin{align}
\rho_{q}&=\sum_{i\in q}\psi_{i}^{\prime q*}\psi_{i}^{q},\label{eq:time-even fields}\\
\tau_{q}&=\sum_{i\in q}\vec{\nabla}\psi_{i}^{\prime q*}\cdot\vec{\nabla}\psi_{i}^{q},\label{eq:time-even fields tau}\\
\vec{J}_{q}&=-i\sum_{i\in q}\psi^{\prime q*}_{i}\left(
\vec{\nabla}\times\vec{\sigma}
\right)\psi_{i}^{q},\label{eq:time-even fields J}
\end{align}
where $\psi_{i}^{q}\equiv\psi_{i}(\vec{r},\sigma,q)$ and $\psi_{i}^{\prime q*}\equiv\psi_{i}^{\prime*}(\vec{r},\sigma,q)$ are the wave functions in Eqs.~(\ref{eq:single-particle states TDHF}) and (\ref{eq:single-particle states TDHF complex conjugate}), respectively. The index $i$ in the sums represents the occupied states of nucleon $q$. The summation of the spin dependence of wave functions can be done automatically in Eqs.~(\ref{eq:time-even fields})-(\ref{eq:time-even fields J}), when the $\phi_{\mu}^{q}$ is composed of both the spin-up state $\chi_{1/2}(\sigma)$ and the spin-down state $\chi_{-1/2}(\sigma)$. In fact, as shown in Eq.~(\ref{eq:single-particle states HF}), the HF single-particle state used in our numerical calculation includes both the spin-up and spin-down states. The contribution of $h^{\mathrm{even}}_{q}$ on the RPA matrices in Eqs.~(\ref{eq:RPA matrix A}) and (\ref{eq:RPA matrix B}) is described by the partial derivatives of Eqs.~(\ref{eq:time-even fields})--(\ref{eq:time-even fields J}). According to Eqs.~(\ref{eq:single-particle states TDHF}) and (\ref{eq:single-particle states TDHF complex conjugate}), the time-even fields in Eqs.~(\ref{eq:time-even fields})--(\ref{eq:time-even fields J}) are expanded by $\eta$. In the limit of $\eta\to0$, partial derivatives such as $\partial/\partial(\eta X_{nj})$ and $\partial/\partial(\eta Y_{nj})$ of the time-even fields are derived analytically, 
\begin{align}
\left(
\frac{\partial \rho_{q}}{\partial (\eta X_{nj}^{q^{\prime}})}\right)_{\eta=0}&=\delta_{qq^{\prime}}\phi^{q*}_{j}\phi_{n}^{q},\label{eq:derivative of X time-even}\\
\left(\frac{\partial \tau_{q}}{\partial (\eta X_{nj}^{q^{\prime}})}\right)_{\eta=0}&=\delta_{qq^{\prime}}\vec{\nabla}\phi^{q*}_{j}\cdot\vec{\nabla}\phi_{n}^{q},\label{eq:derivative of X time-even tau}\\
\left(\frac{\partial (\vec{\nabla}\cdot\vec{J}_{q})}{\partial (\eta X_{nj}^{q^{\prime}})}\right)_{\eta=0}&=\delta_{qq^{\prime}}\left(
-i
\right)\vec{\nabla}\phi^{q*}_{j}\cdot\left(
\vec{\nabla}\times\vec{\sigma}
\right)\phi_{n}^{q},\label{eq:derivative of X time-even J}
\end{align}

\begin{align}
\left(\frac{\partial \rho_{q}}{\partial (\eta Y_{nj}^{q^{\prime}})}\right)_{\eta=0}&=\delta_{qq^{\prime}}\phi^{q*}_{n}\phi_{j}^{q},\label{eq:derivative of Y time-even}\\
\left(\frac{\partial \tau_{q}}{\partial (\eta Y_{nj}^{q^{\prime}})}\right)_{\eta=0}&=\delta_{qq^{\prime}}\vec{\nabla}\phi^{q*}_{n}\cdot\vec{\nabla}\phi_{j}^{q},\label{eq:derivative of Y time-even tau}\\
\left(\frac{\partial (\vec{\nabla}\cdot\vec{J}_{q})}{\partial (\eta Y_{nj}^{q^{\prime}})}\right)_{\eta=0}&=\delta_{qq^{\prime}}\left(
-i
\right)\vec{\nabla}\phi^{q*}_{n}\cdot\left(
\vec{\nabla}\times\vec{\sigma}
\right)\phi_{j}^{q},\label{eq:derivative of Y time-even J}
\end{align}
where $\phi_{j}^{q}$ and $\phi_{n}^{q}$ are occupied and unoccupied single-particle states, respectively. Since the external field $V_{\mathrm{ext}}$ we consider here does not change the isospins of nucleons, Eqs.~(\ref{eq:derivative of X time-even})--(\ref{eq:derivative of Y time-even J}) should be zero when $q\neq q^{\prime}$. The divergence of $\vec{J}_{q}$ has a contribution to the Skyrme potentials \cite{Vautherin:1973zz}. After simple algebra, we can show that Eqs.~(\ref{eq:derivative of Y time-even})--(\ref{eq:derivative of Y time-even J}) are complex conjugates of Eqs.~(\ref{eq:derivative of X time-even})--(\ref{eq:derivative of X time-even J}). Such a property of the complex conjugate in the partial derivatives in the backward amplitudes is also confirmed in the time-odd Hamiltonian. Therefore, the RPA matrix $B$ is given by the complex conjugate of the partial derivative with respect to the forward amplitude,
\begin{equation}
\label{eq:RPA matrix B relation}
\begin{split}
B_{mi,nj}^{q,q^{\prime}}&=\int\mathrm{d}^{3}r\ \phi_{m}^{q*}\left(
\frac{\partial h_{q}}{\partial (\eta X_{nj}^{q^{\prime}})}
\right)_{\eta=0}^{*}\phi_{i}^{q}.
\end{split}
\end{equation}
It is enough to calculate the partial derivative $\partial h_{q}/\partial(\eta X_{nj}^{q^{\prime}})|_{\eta=0}$ only to derive the RPA matrix $B$. Hereafter, we mainly focus on the derivation of the RPA matrix $A$. The Skyrme forces are parameterized by the $t$ and $x$ coefficients \cite{Chabanat:1997un}. We follow the detailed description of Skyrme potentials in the form of $b$ coefficients \cite{Maruhn:2013mpa}. We can calculate the contribution of the time-even potentials to the RPA matrix $A$ by using Eqs.~(\ref{eq:derivative of X time-even})--(\ref{eq:derivative of X time-even J}). For example, the contribution of the effective mass $m^{*}_{q}$ on the RPA matrix $A$ is given by
\begin{align}
\label{eq:derivative of effective mass}
&\int\mathrm{d}^{3}r\ \phi_{m}^{q*}\vec{\nabla}
\cdot\left(\frac{\partial}{\partial (\eta X_{nj}^{q^{\prime}})}
\frac{-\hbar^{2}}{2m^{*}_{q}}\right)_{\eta=0}\vec{\nabla}\phi_{i}^{q} \notag \\
&=-\int\mathrm{d}^{3}r\ \phi_{m}^{q*}\vec{\nabla}
\cdot
\left\{
b_{1}\frac{\partial(\rho_{n}+\rho_{p})}{\partial (\eta X_{nj}^{q^{\prime}})}-b_{1}^{\prime}\frac{\partial\rho_{q}}{\partial (\eta X_{nj}^{q^{\prime}})}
\right\}_{\eta=0}\vec{\nabla}\phi_{i}^{q} \notag\\
&=\left(
b_{1}-\delta_{qq^{\prime}}b_{1}^{\prime}
\right)\int\mathrm{d}^{3}r\ \phi^{q^{\prime}*}_{j}\phi_{n}^{q^{\prime}}\vec{\nabla}\phi_{m}^{q*}\cdot\vec{\nabla}\phi_{i}^{q},
\end{align}
where the term $\vec{\nabla}\phi_{m}^{q*}$ in the third line comes from integration by parts. $b_{1}$ and $b_{1}^{\prime}$ are coefficients in the effective mass \cite{Maruhn:2013mpa}. In the same way, the contribution of the spin-orbit potential is given by
\begin{equation}
\label{eq:derivative of spin-orbit}
\begin{split}
 &\int\mathrm{d}^{3}r\ \phi_{m}^{q*}\left(
 \frac{\partial \vec{W}_{q}}{\partial (\eta X_{nj}^{q^{\prime}})}
 \right)_{\eta=0}\cdot(-i)\left(
    \vec{\nabla}\times\vec{\sigma}
    \right)\phi_{i}^{q}\\
    &=-\left(
b_{4}+\delta_{qq^{\prime}}b_{4}^{\prime}
\right)\int\mathrm{d}^{3}r\ \phi^{q^{\prime}*}_{j}\phi_{n}^{q^{\prime}}\vec{\nabla}\phi_{m}^{q*}\cdot\left(
    -i
    \right)\left(
    \vec{\nabla}\times\vec{\sigma}
    \right)\phi_{i}^{q},
\end{split}
\end{equation}
where the second line is derived from the integration by parts and the property of $\vec{\nabla}\cdot(\vec{\nabla}\times\vec{\sigma})=0$. We calculate the partial derivatives of the central and the Coulomb potentials as Eqs.~(\ref{eq:derivative of effective mass}) and (\ref{eq:derivative of spin-orbit}). Finally, the contribution of $h_{q}^{\mathrm{even}}$ to the RPA matrix $A$ in Eq.~(\ref{eq:RPA matrix A}) is described by
\begin{widetext}
\begin{equation}
\label{eq:partial derivative time-even Hamiltonian}
\begin{split}
\int\mathrm{d}^{3}r\ \phi_{m}^{q*}&\left(
\frac{\partial h^{\mathrm{even}}_{q}}{\partial (\eta X_{nj}^{q^{\prime}})} \right)_{\eta=0}\phi_{i}^{q}=\left(
b_{0}-\delta_{qq^{\prime}}b_{0}^{\prime}
\right)\int\mathrm{d}^{3}r\ 
\phi_{m}^{q*}\phi_{i}^{q}\phi^{q^{\prime}*}_{j}\phi_{n}^{q^{\prime}}\\
&+\left(
b_{1}-\delta_{qq^{\prime}}b_{1}^{\prime}
\right)\int\mathrm{d}^{3}r\ \left(
\phi^{q*}_{m}\phi_{i}^{q}\vec{\nabla}\phi_{j}^{q^{\prime}*}\cdot\vec{\nabla}\phi_{n}^{q^{\prime}}
+\phi^{q^{\prime}*}_{j}\phi_{n}^{q^{\prime}}\vec{\nabla}\phi_{m}^{q*}\cdot\vec{\nabla}\phi_{i}^{q}\right)\\
&-\left(
b_{2}-\delta_{qq^{\prime}}b_{2}^{\prime}
\right)\int\mathrm{d}^{3}r\ \left\{
\phi_{m}^{q*}\phi_{i}^{q}\left(
\phi^{q^{\prime}*}_{j}\nabla^{2}\phi_{n}^{q^{\prime}}+\phi_{n}^{q^{\prime}}\nabla^{2}\phi^{q^{\prime}*}_{j}+2\nabla\phi^{q^{\prime}*}_{j}\cdot\nabla\phi_{n}^{q^{\prime}}
\right)
\right\}\\
&+b_{3}\int\mathrm{d}^{3}r\ \left\{
\frac{(\alpha+2)(\alpha+1)}{3}(\rho_{0})^{\alpha}
\phi_{m}^{q*}\phi_{i}^{q}\phi^{q^{\prime}*}_{j}\phi_{n}^{q^{\prime}}\right\}\\
&-b_{3}^{\prime}\int\mathrm{d}^{3}r\ \left[\left\{
\frac{2\alpha}{3}(\rho_{0})^{\alpha-1}(\rho_{0,q}+\rho_{0,q^{\prime}})
+\frac{2}{3}(\rho_{0})^{\alpha}\delta_{qq^{\prime}}
+\frac{\alpha(\alpha-1)}{3}(\rho_{0})^{\alpha-2}
\sum_{q^{\prime\prime}}(\rho_{0,q^{\prime\prime}})^{2}
\right\}\phi_{m}^{q*}\phi_{i}^{q}\phi^{q^{\prime}*}_{j}\phi_{n}^{q^{\prime}}
\right]\\
&-\left(
b_{4}+\delta_{qq^{\prime}}b_{4}^{\prime}
\right)\int\mathrm{d}^{3}r\ \left\{ 
\phi^{q*}_{m}\phi_{i}^{q}\vec{\nabla}\phi_{j}^{q^{\prime}*}\cdot\left(
    -i
    \right)\left(
    \vec{\nabla}\times\vec{\sigma}
    \right)\phi_{n}^{q^{\prime}}
+\phi^{q^{\prime}*}_{j}\phi_{n}^{q^{\prime}}\vec{\nabla}\phi_{m}^{q*}\cdot\left(
    -i
    \right)\left(
    \vec{\nabla}\times\vec{\sigma}
    \right)\phi_{i}^{q}\right\}\\
&+\delta_{qp}\delta_{q^{\prime}p}\frac{e^{2}}{2}\int\mathrm{d}^{3}r\ \left\{\phi_{m}^{q*}\phi_{i}^{q}\left(
\int\mathrm{d}^{3}r^{\prime}\frac{\phi^{q^{\prime}*}_{j}\phi_{n}^{q^{\prime}}}{|\vec{r}-\vec{r^{\prime}}|}
\right)
-\frac{1}{3}\left(
\frac{3}{\pi}
\right)^{1/3}(\rho_{0,p})^{-2/3}
\phi_{m}^{q*}\phi_{i}^{q}\phi^{q^{\prime}*}_{j}\phi_{n}^{q^{\prime}}\right\},
\end{split}
\end{equation}
\end{widetext}
where $\rho_{0}=\rho_{0,n}+\rho_{0,p}$ is the summation of the density of nucleon, $\rho_{0,q}=\sum_{i\in q}\phi_{i}^{q*}\phi_{i}^{q}\ (q=n,p)$.  $b_{i},b_{i}^{\prime}(i=0,1,2,3,\rm{and}\ 4)$ and $\alpha$ are the parameters in the Skyrme forces \cite{Maruhn:2013mpa}. It is clear from Eq.~(\ref{eq:partial derivative time-even Hamiltonian}) that the residual interaction of $h_{q}^{\mathrm{even}}$ is composed of the single-particle states of the static HF calculation without any forward and backward amplitudes. We show the detailed descriptions of integrands such as $\phi^{q*}_{m}\phi_{i}^{q}$, $\vec{\nabla}\phi^{q*}_{m}\vec{\nabla}\phi_{i}^{q}$, $\phi^{q^{\prime*}}_{j}\nabla^{2}\phi_{n}^{q^{\prime}}$, and $(-i)\vec{\nabla}\phi^{q*}_{m}(\vec{\nabla}\times\vec{\sigma})\phi_{i}^{q}$ in Appendix \ref{sec:detail of integrands}. The first term in the last line of Eq.~(\ref{eq:partial derivative time-even Hamiltonian}) represents the contribution from the direct term of Coulomb potential. We discuss the calculation method for the double spatial integrals of the direct term in Appendix \ref{sec:detail of Coulomb term}.

\subsection{The residual interaction of $h_{q}^{\mathrm{odd}}$}
\label{sec:residula time-odd}

The time-odd Hamiltonian $h_{q}^{\mathrm{odd}}$ itself is a time-even field, but this Hamiltonian is composed of time-odd fields such as the spin density $\vec{s}_{q}$ and the current density $\vec{j}_{q}$ \cite{Engel:1975zz,Maruhn:2013mpa}. The time-odd fields of nucleon $q$ are described by
\begin{align}
\vec{s}_{q}&=\sum_{i\in q}\psi^{\prime q*}_{i}\vec{\sigma}\psi_{i}^{q},\label{eq:time-odd fields}\\
\vec{j}_{q}&=\frac{1}{2i}\sum_{i\in q}\left\{
\psi^{\prime q*}_{i}\vec{\nabla}\psi_{i}^{q}-\psi_{i}^{q}\vec{\nabla}\psi_{i}^{\prime q*}
\right\}_{.}\label{eq:time-odd fields j}
\end{align}
In the limit of $\eta\to0$, the time-odd fields in the equations above are converged to zero when the time-reversal symmetry is satisfied in the HF calculation. As done in the time-even fields, We can expand Eqs.~(\ref{eq:time-odd fields}) and (\ref{eq:time-odd fields j}) by using Eqs.~(\ref{eq:single-particle states TDHF}) and (\ref{eq:single-particle states TDHF complex conjugate}). Then, we obtain the partial derivatives of these time-odd fields,
\begin{align}
\left(
\frac{\partial \vec{s}_{q}}{\partial (\eta X_{nj}^{q^{\prime}})}\right)_{\eta=0}&=\delta_{qq^{\prime}}\phi^{q^{\prime}*}_{j}\vec{\sigma}\phi_{n}^{q^{\prime}},\label{eq:derivative of X time-odd}\\
\left(
\frac{\partial \vec{j}_{q}}{\partial (\eta X_{nj}^{q^{\prime}})}\right)_{\eta=0}&=\delta_{qq^{\prime}}\frac{1}{2i}\left(
\phi^{q^{\prime}*}_{j}\vec{\nabla}\phi_{n}^{q^{\prime}}-\phi^{q^{\prime}}_{n}\vec{\nabla}\phi_{j}^{q^{\prime}*}
\right),\label{eq:derivative of X time-odd j}
\end{align}

\begin{align}
\left(
\frac{\partial \vec{s}_{q}}{\partial (\eta Y_{nj}^{q^{\prime}})}\right)_{\eta=0}&=\delta_{qq^{\prime}}\phi^{q^{\prime}*}_{n}\vec{\sigma}\phi_{j}^{q^{\prime}},\label{eq:derivative of Y time-odd}\\
\left(
\frac{\partial \vec{j}_{q}}{\partial (\eta Y_{nj}^{q^{\prime}})}\right)_{\eta=0}&=\delta_{qq^{\prime}}\frac{1}{2i}\left(
\phi^{q^{\prime}*}_{n}\vec{\nabla}\phi_{j}^{q^{\prime}}-\phi^{q^{\prime}}_{j}\vec{\nabla}\phi_{n}^{q^{\prime}*}
\right),\label{eq:derivative of Y time-odd j}
\end{align}
where Eqs.~(\ref{eq:derivative of Y time-odd}) and (\ref{eq:derivative of Y time-odd j}) are the complex conjugates of Eqs.~(\ref{eq:derivative of X time-odd}) and (\ref{eq:derivative of X time-odd j}). As discussed in Sec.\ref{sec:residula time-even}, such the complex conjugate relations lead to the property of the RPA matrix $B$ in Eq.~(\ref{eq:RPA matrix B relation}). The time-odd potentials such as $\vec{S}_{q}$ and $\vec{A}_{q}$ in $h_{q}^{\mathrm{odd}}$ are composed of the time-odd fields in Eqs.~(\ref{eq:time-odd fields}) and (\ref{eq:time-odd fields j}). The detail of these potentials is shown in Ref.~\cite{Maruhn:2013mpa}. By using Eqs.~(\ref{eq:TDHF Hamiltonian odd}), (\ref{eq:derivative of X time-odd}), and (\ref{eq:derivative of X time-odd j}), the partial derivative of $h_{q}^{\mathrm{odd}}$ in the RPA matrix $A$ is derived;
\begin{widetext}
\begin{align}
\label{eq:partial derivative time-odd Hamiltonian}
    &\int\mathrm{d}^{3}r\ \phi_{m}^{q*}\left(
\frac{\partial h^{\mathrm{odd}}_{q}}{\partial (\eta X_{nj}^{q^{\prime}})} \right)_{\eta=0}\phi_{i}^{q} \notag\\
=&\int\mathrm{d}^{3}r\ 
\left\{
\frac{\partial \vec{S}_{q}}{\partial (\eta X_{nj}^{q^{\prime}})}\cdot(\phi^{q*}_{m}\vec{\sigma}\phi_{i}^{q})
+\frac{\partial \vec{A}_{q}}{\partial (\eta X_{nj}^{q^{\prime}})}\cdot\frac{1}{2i}(\phi^{q*}_{m}\vec{\nabla}\phi_{i}^{q}-\phi^{q}_{i}\vec{\nabla}\phi_{m}^{q^{*}})
\right\}_{\eta=0} \notag\\
=&-2(b_{1}-\delta_{qq^{\prime}}b_{1}^{\prime})\int\mathrm{d}^{3}r\ \left\{
\frac{1}{2i}(\phi^{q*}_{m}\vec{\nabla}\phi_{i}^{q}-\phi^{q}_{i}\vec{\nabla}\phi_{m}^{q^{*}})\cdot\frac{1}{2i}(\phi^{q^{\prime}*}_{j}\vec{\nabla}\phi_{n}^{q^{\prime}}-\phi^{q^{\prime}}_{n}\vec{\nabla}\phi_{j}^{q^{\prime}*})
\right\} \notag\\
&-(b_{4}+\delta_{qq^{\prime}}b_{4}^{\prime})\int\mathrm{d}^{3}r\ \left\{
\frac{1}{2i}(\phi^{q*}_{m}\vec{\nabla}\phi_{i}^{q}-\phi^{q}_{i}\vec{\nabla}\phi_{m}^{q^{*}})\cdot\vec{\nabla}\times(\phi^{q^{\prime}*}_{j}\vec{\sigma}\phi_{n}^{q^{\prime}})
+\frac{1}{2i}(\phi^{q^{\prime}*}_{j}\vec{\nabla}\phi_{n}^{q^{\prime}}-\phi^{q^{\prime}}_{n}\vec{\nabla}\phi_{j}^{q^{\prime}*})\cdot\vec{\nabla}\times(\phi^{q*}_{m}\vec{\sigma}\phi_{i}^{q})
\right\},
\end{align}
\end{widetext}
where the integration by parts is done in the term of $\vec{A}_{q}$ from the first line to the second one. The detailed descriptions of the integrands in our numerical calculation are given in Appendix \ref{sec:detail of integrands}.

Finally, the matrices in Eqs.~(\ref{eq:RPA matrix A}) and (\ref{eq:RPA matrix B}) are derived from Eqs.~(\ref{eq:RPA matrix B relation}), (\ref{eq:partial derivative time-even Hamiltonian}), and (\ref{eq:partial derivative time-odd Hamiltonian}). From these equations, we can also confirm that these matrices are Hermitian ($A_{mi,nj}^{q,q^{\prime}}=A_{nj,mi}^{q^{\prime},q*}$) and symmetric ($B_{mi,nj}^{q,q^{\prime}}=B_{nj,mi}^{q^{\prime},q}$), which is consistent with the general property of the RPA matrices \cite{ring2004nuclear}. 
The RPA matrices do not include $X_{mi}^{q}$, and $Y_{mi}^{q}$, so that the forward and backward amplitudes are obtained by solving Eq.~(\ref{eq:RPAeq}) without the need for a iterative procedure.

Note that all the RPA equations given above are in the matrix form. The same linearizatoin could be also applied to the mixed representation~\cite{Inakura:2009vs}.

Extension of non-iterative FAM-RPA to FAM-QRPA is straightforward by employing the same technique described by Avogadro and Nakatsukasa~\cite{Avogadro:2011gd}. This certainly increases the number of particle-hole configurations. However, they are important only for the states with low excitation energies near the Fermi surface. We limit ourselves to FAM-RPA in this paper.

\subsection{External fields for E$1$ and M$1$ transitions}

The calculated results of FAM-RPA can be compared with experimental data of the photoabsorption cross section. Here, we use a complex frequency, $\omega=E+i\gamma/2$. $E$ and $\gamma$ correspond to the energy of the incoming photon and the Lorentzian width. This $\gamma$ parameter characterizes the width of the photoabsorption cross section. The transition strength is described by the forward and backward amplitudes \cite{Nakatsukasa:2007qj},
\begin{align}
\label{eq:strength function}
\frac{\mathrm{d}B(E;V_{\mathrm{ext}})}{\mathrm{d}E}
&=-\frac{1}{\pi}\mathrm{Im}\sum_{q}\sum_{m,i\in q}\left(
f^{q*}_{mi}X_{mi}^{q}+f_{im}^{q*}Y_{mi}^{q}
\right).
\end{align}
As the external field, we consider the electric and magnetic dipole operators that induce E$1$ and M$1$ transitions. The electric dipole operator is written as the spherical harmonics of neutrons and protons,
\begin{equation}
\label{eq:electric dipole operator}
\begin{split}
D_{K}=\sum_{i=1}^{A}e_{\mathrm{eff}}^{(i)}r_{i}Y_{1K}(\theta_{i},\varphi_{i})\ \ (K=0,\pm1),
\end{split}
\end{equation}
where $e_{\mathrm{eff}}^{(i)}\equiv-eZ/A\ (eN/A)$ for neutrons (protons). The $(r_{i},\theta_{i},\varphi_{i})$ represents the spherical coordinate of nucleon $i$. The photoabsorption cross section of E$1$ transition is given by \cite{Inakura:2009vs,ring2004nuclear},
\begin{equation}
\label{eq:cross section E1}
\sigma_{\mathrm{abs}}(E;\mathrm{E}1)=\frac{16\pi^{3}}{9\hbar c}E\sum_{K=0,\pm1}\frac{\mathrm{d}B(E;D_{K})}{\mathrm{d}E},
\end{equation}
where we impose $V_{\mathrm{ext}}=\sum_{K=0,\pm1}D_{K}$ on Eq.~(\ref{eq:strength function}). In the case of M$1$ transition, the magnetic dipole operator is represented by operators of a spin part $\vec{\sigma}_{i}/2$ and an orbital part $\vec{l}_{i}=-i(\vec{r}_{i}\times\vec{\nabla}_{i})$ of nucleon $i$, 
\begin{equation}
\begin{split}
\label{eq:magnetic dipole operator}
M_{K}=\mu_{N}\sum_{i=1}^{A}\left(
g_{s}^{(i)}\frac{\vec{\sigma}_{i}}{2}+g_{l}^{(i)}\vec{l}_{i}
\right)\cdot\vec{\nabla}\left(
r_{i}Y_{1K}(\theta_{i},\varphi_{i})
\right)&\\
(K=0,\pm1)&,
\end{split}
\end{equation}
where $\mu_{N}$ is the nuclear magneton. The $g$ factors are $g_{s}^{(i)}=-3.826(5.586)$ and $g_{l}^{(i)}=0(1)$ for neutrons (protons). Similar to  the case of the E$1$ transition, the cross section of M$1$ transition can be expressed as
\begin{equation}
\label{eq:cross section M1}
\sigma_{\mathrm{abs}}(E;\mathrm{M}1)=\frac{16\pi^{3}}{9\hbar c}E\sum_{K=0,\pm1}\frac{\mathrm{d}B(E;M_{K})}{\mathrm{d}E}.
\end{equation}
We note that, for the numerical calculation, we employ the single-particle states labeled in the cylindrical coordinate. Therefore, the spherical coordinates of nucleons in Eqs.~(\ref{eq:electric dipole operator}) and (\ref{eq:magnetic dipole operator}) should be transformed to the cylindrical coordinate when the coefficients in Eq.~(\ref{eq:external field f}) are calculated (see the detail in Appendix \ref{sec:detail of the coefficient of external fields}).


\section{Results and Discussions}
\label{sec:Results and Discussions}

\subsection{E1 transition}
\label{sec:E1 transition}
\subsubsection{Benchmark calculation}

The photoabsorption cross sections of both E$1$ and M$1$ transitions are calculated based on the RPA equations as given in Sec.\ref{sec:theory}. The RPA calculations use the single-particle states of the HF+BCS calculation \cite{Bonneau:2007dc}. In the HF+BCS calculation, we employ the parameterization of SLy4 \cite{Chabanat:1997un} for Skyrme interactions. The detailed description of the single-particle state in the cylindrical coordinate is given in Appendix \ref{sec:detail of integrands}. We focus on even-even nuclei whose HF equations satisfy the time-reversal symmetry. In the case of axially deformed nuclei, we employ the deformation parameters in Table $1$ of Ref.~\cite{Moller:2015fba} to obtain the initial single-particle potentials for the HF$+$BCS calculation to speed up the HF convergence. 

We apply the technique discussed above to a wide mass range of nuclei in order to calculate the cross section of the E$1$ transition. We solve Eq.~(\ref{eq:RPAeq}) for the photon energies starting from $E=0.5$ MeV at every $250$ keV. The resonance energy of GDR becomes lower as the mass number of the nuclei increases \cite{Berman:1975tt}, so that we perform the RPA calculation for light nuclei ($A\leq150$) and heavy nuclei ($A>150$) up to $E=40$ MeV and $30$ MeV, respectively. Since the Lorentzian width $\gamma$ is a free parameter in the FAM calculation, we employ the recommended experimental GDR width $\Gamma$ within the Standard Lorentzian approach in Table III of Ref.~\cite{Kawano:2019tsn}. 

Here, we determine the size of the configuration space of the unoccupied states by imposing a cutoff energy $E_{\mathrm{cut}}$ measured from the Fermi energy surface. We consider the unoccupied state of nucleon $q$ whose single-particle energy satisfies: $\epsilon_{m}<E_{\mathrm{cut}}+\epsilon_{i,q,\mathrm{max}}$, where the $\epsilon_{i,q,\mathrm{max}}$ is the maximum energy of the occupied states of neutron ($q=n$) or proton ($q=p$). 
\begin{figure}[t]
\includegraphics[width=1\linewidth]{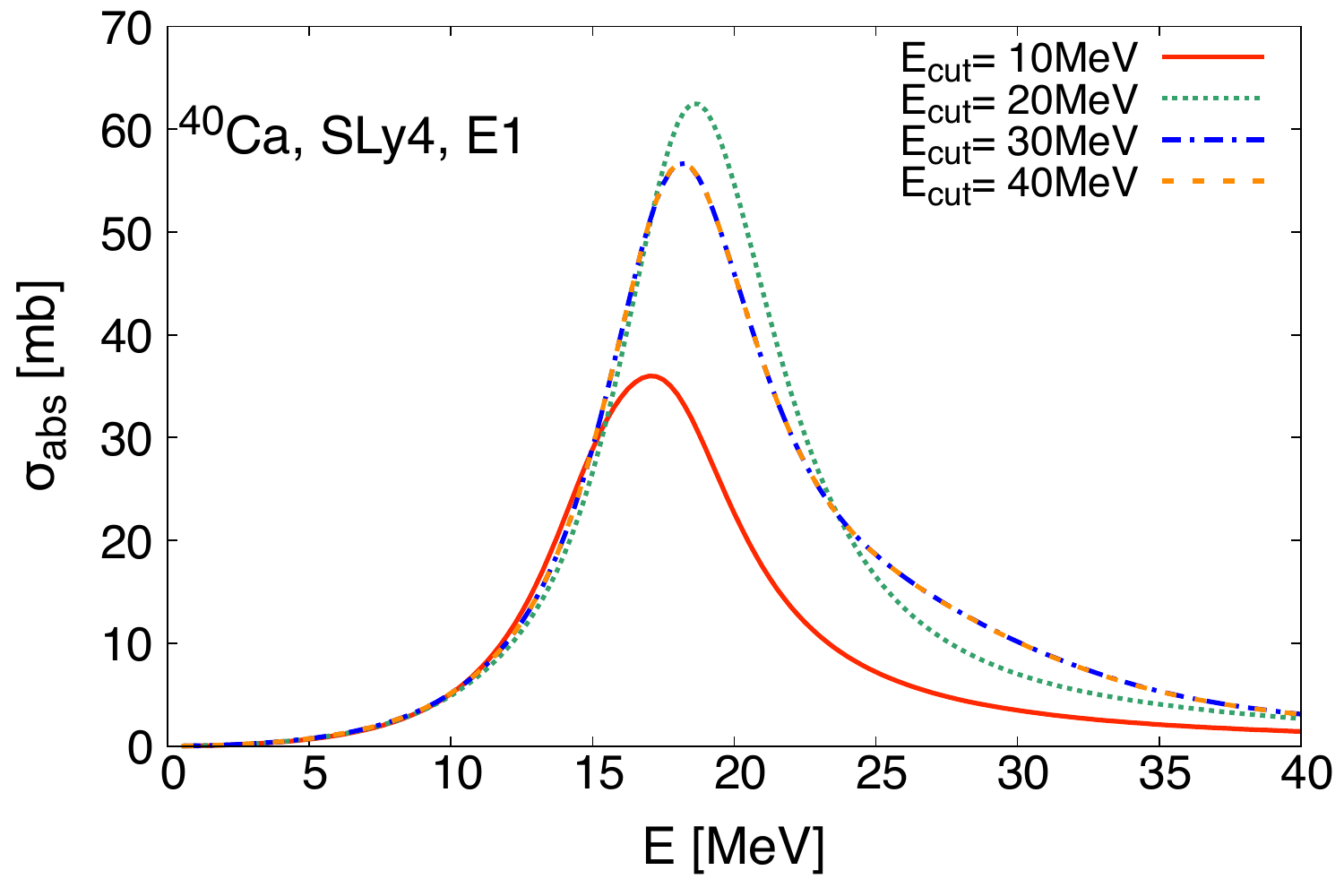}\\
\caption{
The calculated GDR cross sections in Eq.~(\ref{eq:cross section E1}) for $^{40}\mathrm{Ca}$ with different energy cutoff $E_{\mathrm{cut}}$. The result of 30~MeV (dash-dotted line) is almost identical to that of 40~MeV (dashed line).}
\label{fig:energy_cutoff_Ca40}
\end{figure}
Figure~\ref{fig:energy_cutoff_Ca40} shows the sensitivity of the $E_{\mathrm{cut}}$ to the E$1$ cross section of $^{40}\mathrm{Ca}$. The Lorentzian width is $\gamma=6.2$ MeV in $K=0,\pm1$. As shown by the almost identical dash-dotted and dashed lines, we can safely say that the calculation converges when $E_{\rm cut} > 30$~MeV. When we increase $E_{\rm cut}$, the tail on the higher side of GDR becomes slightly larger, because more energetically higher unoccupied states are involved, although these transitions are weak. The energy-weighted sum rule $m_{1}$ \cite{Oishi:2015lph} can be estimated from the energy integration of the photoabsorption cross section. In the case of $E_{\mathrm{cut}}=30$ MeV (dash-dotted line), we obtain $m_{1}=148.5\ e^{2}\mathrm{fm}^{2}\mathrm{MeV}$. The energy-weighted sum rule exhausts $99\%$ of the Thomas-Reiche-Kuhn (TRK) sum rule \cite{ring2004nuclear}. The value of $m_{1}$ in our calculation should exceed that of the TRK sum rule if the cross section is integrated up to the higher energy region ($E>40$ MeV). Such excess of $m_{1}$ reflects the contribution from many-body interactions and depends on the model of the Skyrme forces employed.

In the case of heavy nuclei ($A>150$), the convergence for $E_{\rm cut}$ is faster, since the single-particle state density becomes higher. We adopt the cutoff energy, $E_{\mathrm{cut}}=30$ MeV for $A\leq150$ and $20$ MeV for $A>150$.

\subsubsection{Spherical nuclei}
\label{sec:Spherical nuclei E1}
\begin{figure*}
\subfigure{%
    \includegraphics[clip, width=1\columnwidth]{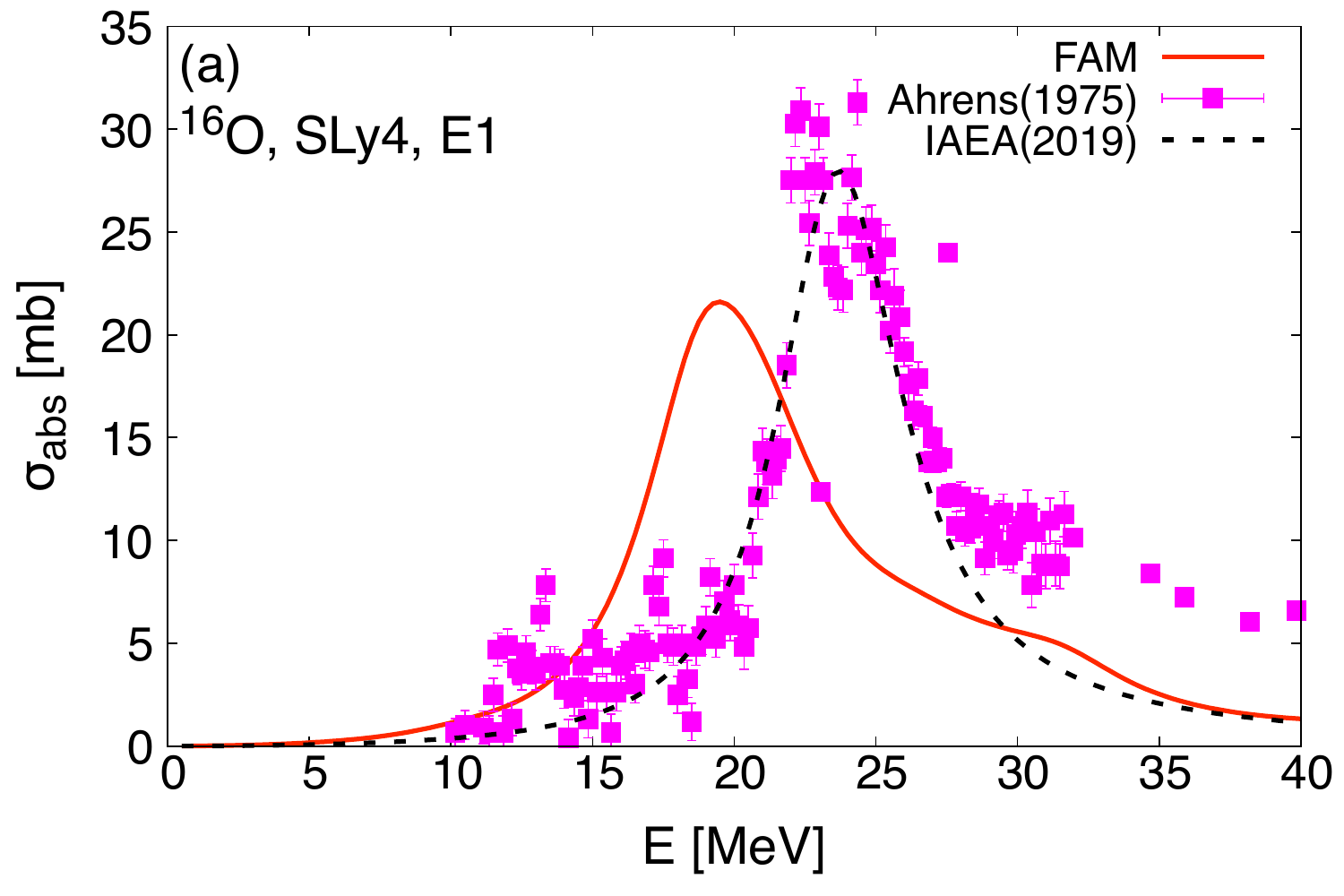}}%
\subfigure{%
    \includegraphics[clip, width=1\columnwidth]{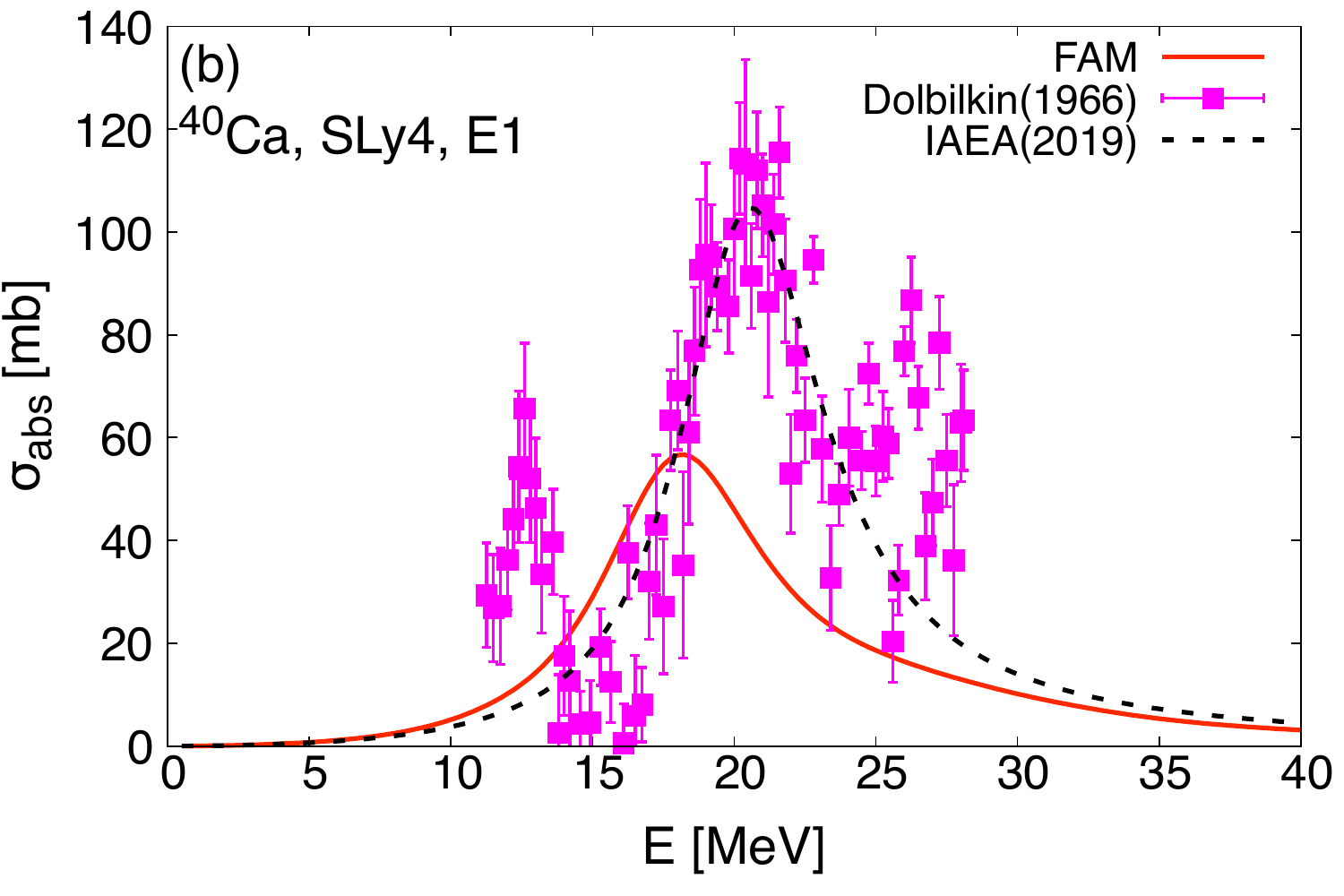}}%
\end{figure*}
\begin{figure*}
\subfigure{%
    \includegraphics[clip, width=1\columnwidth]{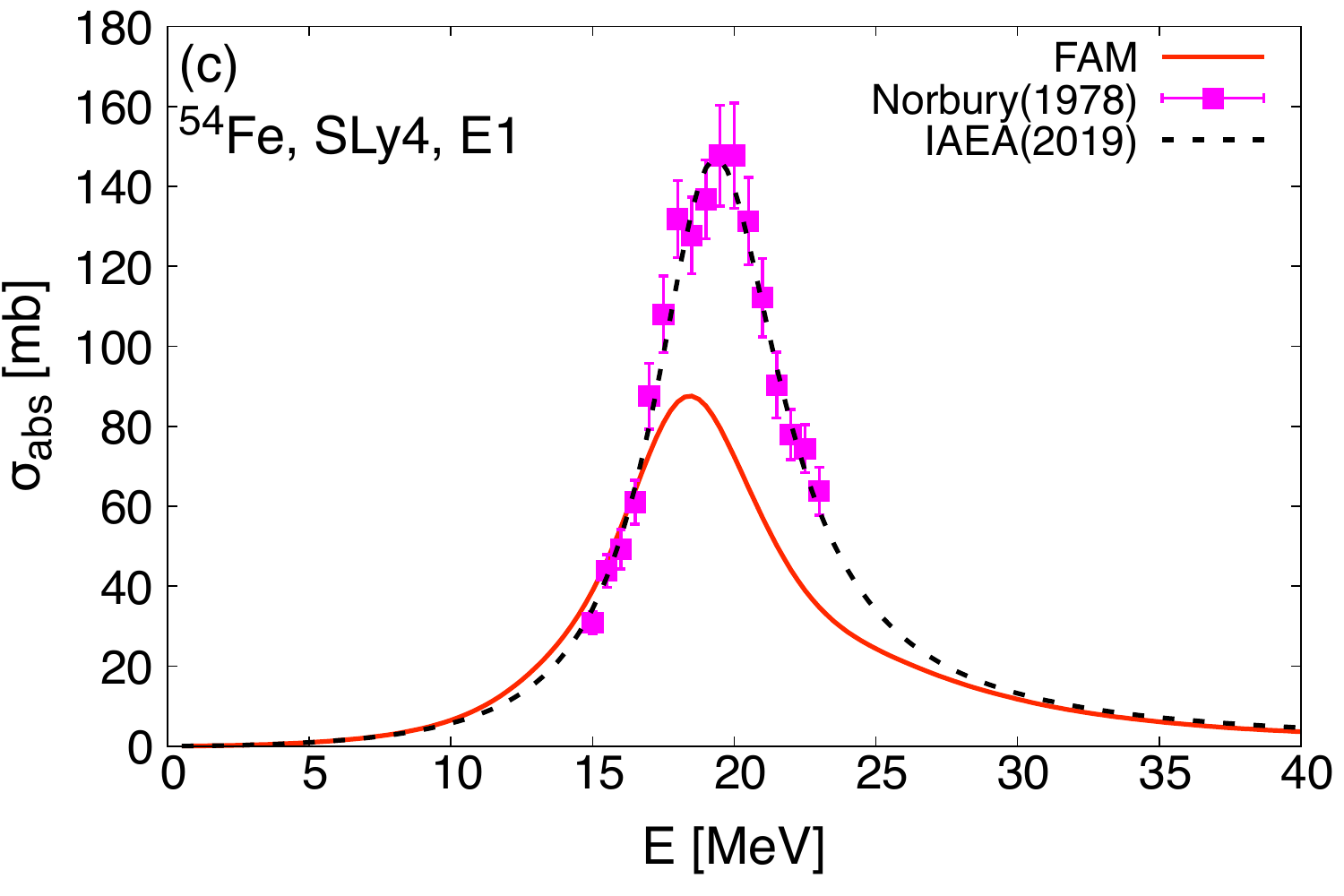}}%
\subfigure{%
    \includegraphics[clip, width=1\columnwidth]{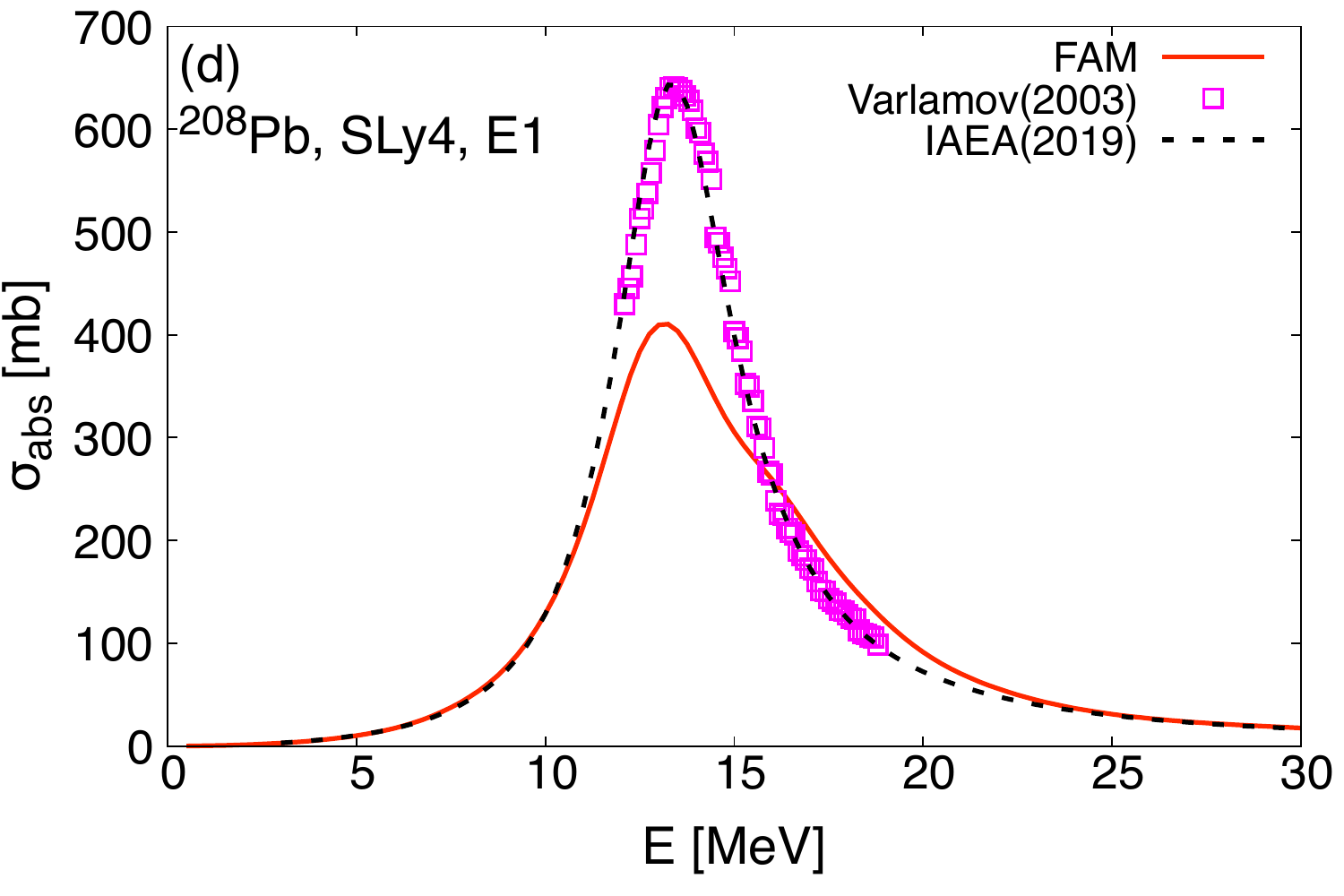}}%
\caption{The GDR cross sections for spherical nuclei, (a) $^{16}\mathrm{O}$, (b) $^{40}\mathrm{Ca}$, (c) $^{54}\mathrm{Fe}$, and (d) $^{208}\mathrm{Pb}$. The solid and dashed lines show the results of Eq.~(\ref{eq:cross section E1}) and the evaluated $\sigma_{\mathrm{GDR}}$ in Eq.~(\ref{eq:Lorenzian fit}), respectively. The symbols represent both the reported and evaluated experimental data~\cite{1975Ahr,1966Dol,1978Nor,2003Var} in EXFOR~\cite{Plujko:2018uum}.}
\label{fig:E1 spherical nuclei}
\end{figure*}

\begin{figure}[t]
\includegraphics[width=1\linewidth]{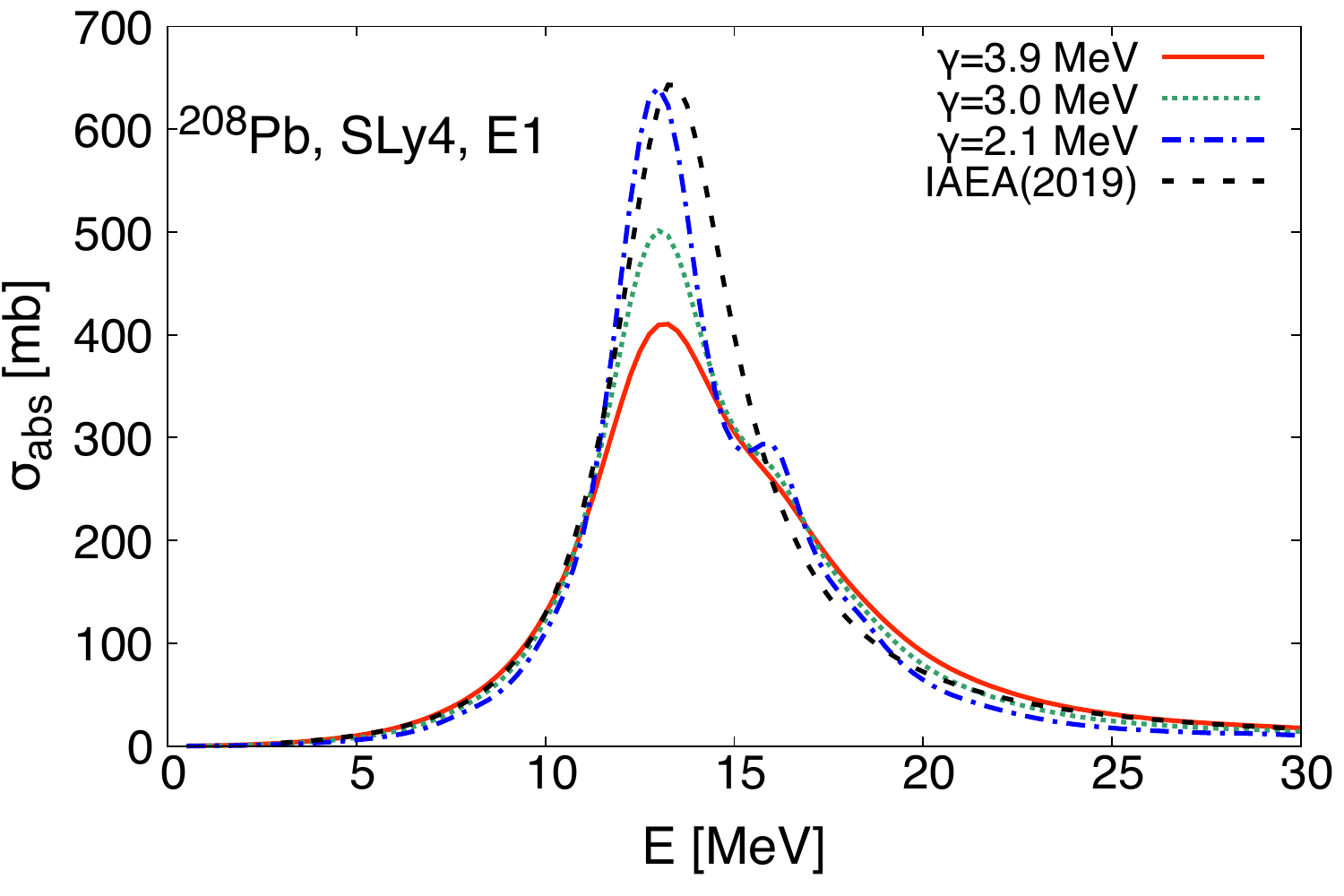}\\
\caption{
The calculated GDR cross sections for $^{208}\mathrm{Pb}$ with different Lorentzian width $\gamma$ (solid, dotted, and dash-dotted lines) and the evaluated $\sigma_{\mathrm{GDR}}$ in Eq.~(\ref{eq:Lorenzian fit}) (dashed line). The result of $\gamma=3.9$ MeV (solid line) corresponds to the FAM result in Fig.~\ref{fig:E1 spherical nuclei}(d).
}
\label{fig:gamma_Pb208}
\end{figure}

Figure \ref{fig:E1 spherical nuclei} shows the cross sections of E$1$ transitions for spherical nuclei such as $^{16}\mathrm{O},$$^{40}\mathrm{Ca},$$^{54}\mathrm{Fe},$ and $^{208}\mathrm{Pb}$. In the case of spherical nucleus, the transition strength, $\mathrm{d}B(E,D_{K})/\mathrm{d}E\ (K=0,\pm1)$ is independent of the value of $K$. The solid lines show the numerical results of Eq.~(\ref{eq:cross section E1}) in our FAM calculations. The symbols are experimental data or evaluated experimental data of the photoabsorption cross section in EXFOR \cite{Plujko:2018uum}. The error bar is not shown in the evaluated data of $^{208}\mathrm{Pb}$ \cite{2003Var} in Fig.~\ref{fig:E1 spherical nuclei}(d). The experimental GDR data are often represented by the Lorentzian distribution \cite{Berman:1975tt},
\begin{equation}
\label{eq:Lorenzian fit}
    \sigma_{\mathrm{GDR}}(E)=\frac{\sigma_{R}}{1+[(E^{2}-E_{R}^{2})^{2}/E^{2}\Gamma_{R}^{2}]},
\end{equation}
where $\sigma_{R}, \Gamma_{R},$ and $E_{R}$ are the peak cross section, the full width at half-maximum, and the resonance energy, respectively. The dashed lines in Fig. \ref{fig:E1 spherical nuclei} show the Lorentzian parameterized experimental GDR compiled at IAEA \cite{Kawano:2019tsn}.

For the light nuclei, the GDR resonance energy $E_{R}$ in the FAM calculation tends to be smaller than that of experimental data. In the case of $^{16}\mathrm{O}$ (Fig.~\ref{fig:E1 spherical nuclei}(a)), $E_{R}$ is lower by $4.3$ MeV. As shown in Fig.~\ref{fig:E1 spherical nuclei}(b), the deviation of $E_{R}$ is $2.3$ MeV for $^{40}\mathrm{Ca}$. The discrepancy of the GDR peak is more noticeable in light nuclei. Such disagreement is also seen in the previous RPA calculations \cite{Inakura:2009vs,Erler:2010fe}. The experimental data of $E_{R}$ can be explained by the superposition of the Goldhaber-Teller mode that produces the dependence $E_{R}\propto A^{-1/6}$ and the Steinwedel-Jensen mode that produces $E_{R}\propto A^{-1/3}$ \cite{Berman:1975tt,Myers:1977zz}. The $A$ dependence of $E_{R}$ in experiments can be fitted by $A^{-1/6}$ for light nuclei. The RPA description underestimates the contribution of the surface mode (Goldhaber-Teller mode), which may imply an insufficient isovector surface energy in the present Skyrme forces \cite{Erler:2010fe}.

As shown in Figs.~\ref{fig:E1 spherical nuclei}(c) and \ref{fig:E1 spherical nuclei}(d), the deviations between the resonance energy of FAM and that of $\sigma_{\mathrm{GDR}}$ are $0.85$ MeV and $0.37$ MeV in $^{54}\mathrm{Fe}$ and $^{208}\mathrm{Pb}$, respectively. Therefore,  the discrepancy between the FAM calculations (solid lines) and the evaluated $\sigma_{\mathrm{GDR}}$ (dashed lines) becomes smaller for the heavy nuclei. It seems such good reproduction of the resonance energy for the heavy nuclei is a common property of RPA calculations \cite{Erler:2010fe}. 
The peak cross sections in the FAM calculations are much smaller than those of evaluated data, $\sigma_{R}$. In the FAM calculations, we assume that the Lorentzian width $\gamma$ is equal to the width, $\Gamma_{R}$ of recommended experimental GDR parameters in Ref.~\cite{Kawano:2019tsn}. This assumption overestimates the value of $\gamma$ and reduces the peak cross section in FAM calculations. The total width of GDR is given by the sum of three widths, $\Gamma_{R}=\Delta \Gamma+\Gamma^{\uparrow}+\Gamma^{\downarrow}$ where $\Delta \Gamma,\ \Gamma^{\uparrow}$, and $\Gamma^{\downarrow}$ are the Landau damping, the escape width, and the spreading width, respectively \cite{harakeh2001giant,Wambach:1988wid,Speth:1981gdr}. The spreading width $\Gamma^{\downarrow}$ represents the couplings of the $1$p-$1$h states to more complex numerous configurations such as $2$p-$2$h, and $3$p-$3$h states \cite{Speth:1981gdr}. The Lorentzian width $\gamma$ can be regarded as $\Gamma^{\downarrow}$ \cite{Inakura:2009vs,Yoshida:2010zu}. Therefore, the total GDR width in FAM calculation, $\Gamma_{R}^{\mathrm{FAM}}$ should be larger than the $\Gamma_{R}$ evaluated based on the experimental data when we assume $\gamma=\Gamma_{R}$. In the case of Fig.~\ref{fig:E1 spherical nuclei}(d)
, for example, we can estimate the $\Gamma_{R}^{\mathrm{FAM}}$ by fitting the numerical result (solid line) with Eq.~(\ref{eq:Lorenzian fit}). We confirm that the $\Gamma_{R}^{\mathrm{FAM}}=5.9$ MeV is larger than $\gamma=3.9$ MeV. Then, the damping width in the RPA is given by $\Delta\Gamma+\Gamma^{\uparrow}=2.0$ MeV. In Fig. \ref{fig:gamma_Pb208}, we compare the FAM results for $^{208}\mathrm{Pb}$ with different $\gamma$ values. The $\gamma = 3.9$ MeV case in Fig. \ref{fig:gamma_Pb208} is identical to Fig.~\ref{fig:E1 spherical nuclei}(d). As $\gamma$ decreases, the peak cross section (GDR width) increases (decreases). The energy-weighted sum rule is almost constant regardless the value of $\gamma$. In Fig.~\ref{fig:gamma_Pb208}, the peak cross section of evaluated experimental data (dash line) is well reproduced by the FAM calculation of the $\gamma=2.1$ MeV. 
The value of the $\Delta\Gamma+\Gamma^{\uparrow}$ might depend on the types of Skyrme forces because the sum rule enhancement factor, $\kappa$, is different depending on the type of Skyrme forces \cite{Bender:2003jk,Klupfel:2008af}. A shoulder structure appears near $16$ MeV, which is on the right-hand side of the resonance energy, when we employ the narrower Lorentzian width. The right shoulder of $^{208}\mathrm{Pb}$ is also confirmed in the previous RPA calculation of the isovector strength function \cite{Nesterenko:2002vq}. Such fragmentation of GDR might be universal for GDR of heavy nuclei irrespective of the nuclear shape \cite{Nesterenko:2006qr}.

\subsubsection{Deformed nuclei}
\label{sec:Deformed nuclei E1}

\begin{figure}[htbp]
\includegraphics[width=1\linewidth]{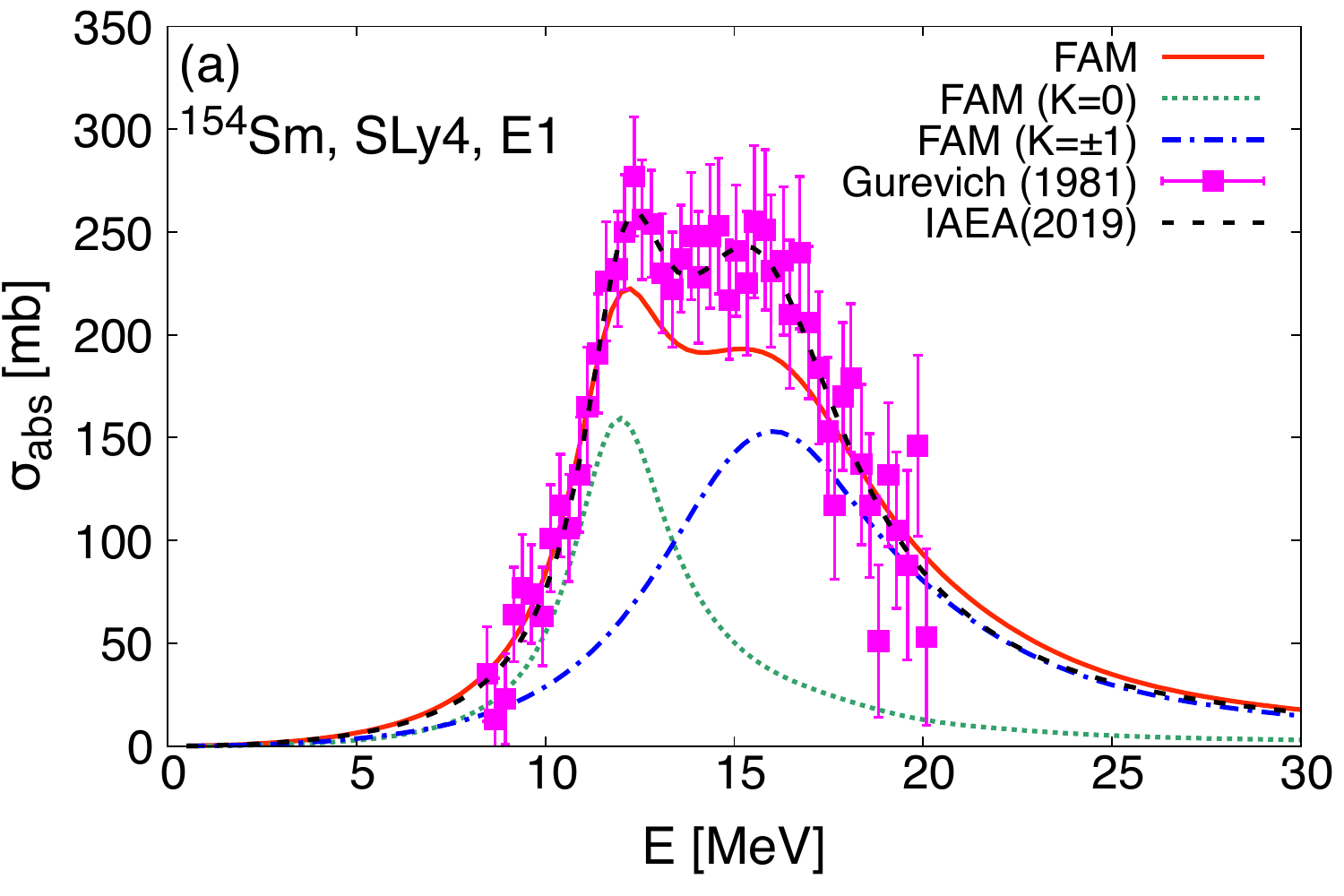}\\
\includegraphics[width=1\linewidth]{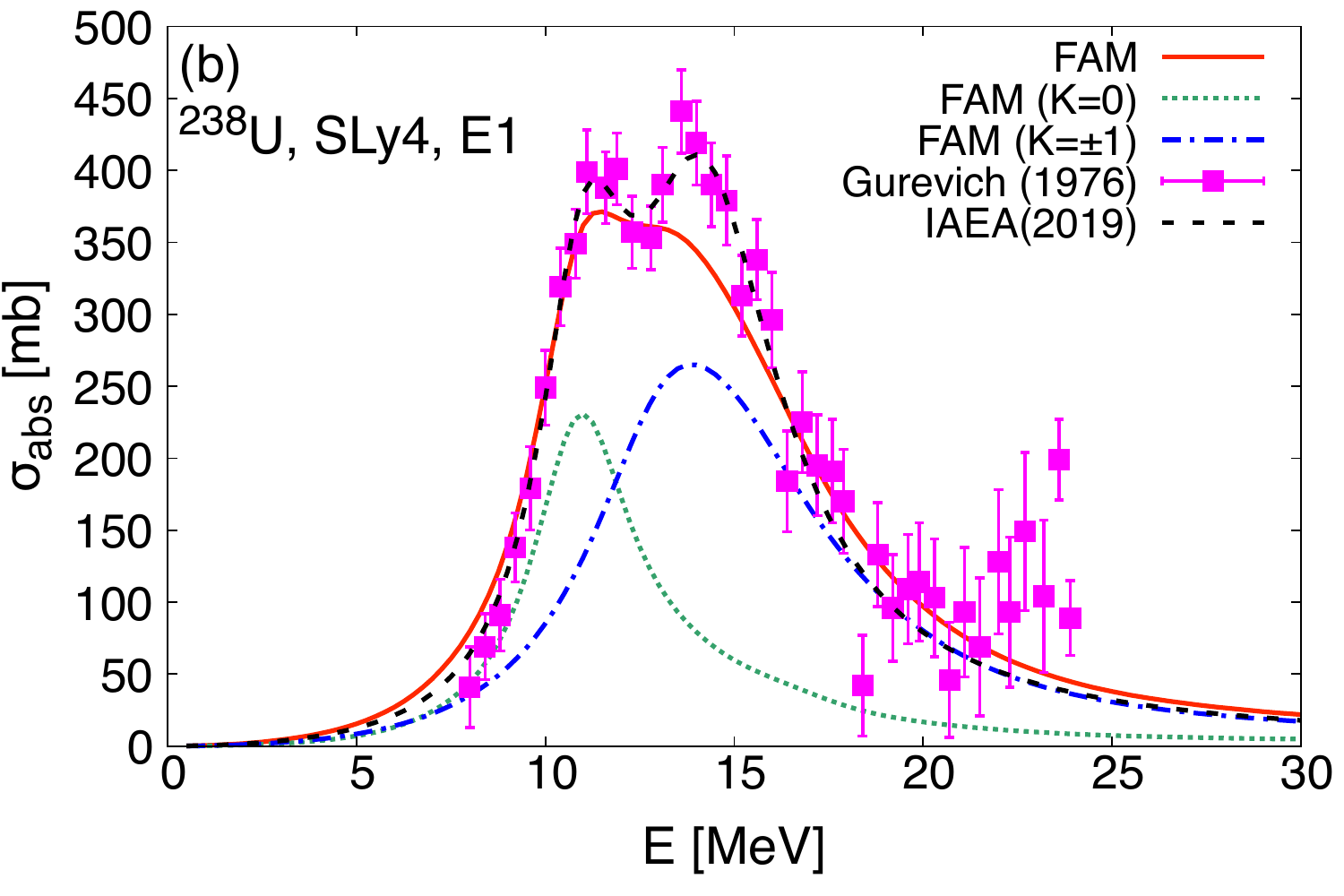}
\caption{
The GDR cross sections for deformed nuclei, (a) $^{154}\mathrm{Sm}$ and (b) $^{238}\mathrm{U}$. The partial contributions from $\mathrm{d}B(E,D_{0})/\mathrm{d}E$ and $\sum_{K=\pm1}\mathrm{d}B(E,D_{K})/\mathrm{d}E$ in Eq.~(\ref{eq:cross section E1}) are shown by the dotted and dash-dotted lines. The symbols represent the experimental data~\cite{Gurevich:1981sm154,Gurevich:1976u238}.
}
\label{fig:E1 deformed nuclei}
\end{figure}

It is well known experimentally that the GDR peak splits into two for the statically deformed nuclei \cite{Berman:1975tt}, because the degeneracy for different $K$ values of 0, $\pm 1$ is removed. As shown in Fig.~\ref{fig:E1 deformed nuclei}, FAM reproduces the split of GDR for well-deformed nuclei such as $^{154}\mathrm{Sm}$ and $^{238}\mathrm{U}$. The resonance energy of $K=0$ is smaller than that of $K=\pm1$ because of the longer wavelength along the axis of symmetry ($z$-axis), when the shape is prolate \cite{Speth:1981gdr}. Because the pair correlations are no longer negligible for deformed nuclei, QRPA is commonly adopted to calculate GDRs of deformed nuclei \cite{Yoshida:2010zu,Oishi:2015lph}. Although the pairing correlation is included in the single-particle states of HF+BCS, the particle and hole are chosen as fully unoccupied and occupied states above and below the Fermi surface given by BCS in our non-iterative FAM-RPA. This approximation results in some uncertainties in the transitions with very small excitation energies. Nevertheless, as shown in Fig.~\ref{fig:E1 deformed nuclei} by the solid lines, the FAM-RPA calculations for $\sigma_{\mathrm{abs}}(E,E_{1})$ nicely reproduce the experimental data. This is because the E$1$ operator in Eq.~(\ref{eq:electric dipole operator}) changes the parities of single-particle states, hence transitions within the same shells are strongly suppressed. These small energy transitions do not contribute to the main part of E1, even if we extend our model to FAM-QRPA, which is definitely planned.

In Fig.~\ref{fig:E1 deformed nuclei}(a), our calculation predicts somewhat lower cross sections than the evaluated Lorentzian distribution, which is also seen in the QRPA calculation for $^{154}\mathrm{Sm}$ where SLy$4$ is employed \cite{Yoshida:2010zu}. In both $^{154}\mathrm{Sm}$ and $^{238}\mathrm{U}$, the second peak cross sections of FAM results are lower than the experimental data. This systematic disagreement occurs due to the Lorentzian width $\gamma$ as we discussed before. The agreement on the second peak and the higher energy tail ($E>16$ MeV) can be improved when a slightly smaller $\gamma$ is used for $K=\pm1$.


\subsection{M1 transition}

\begin{figure}[htbp]
\includegraphics[width=1\linewidth]{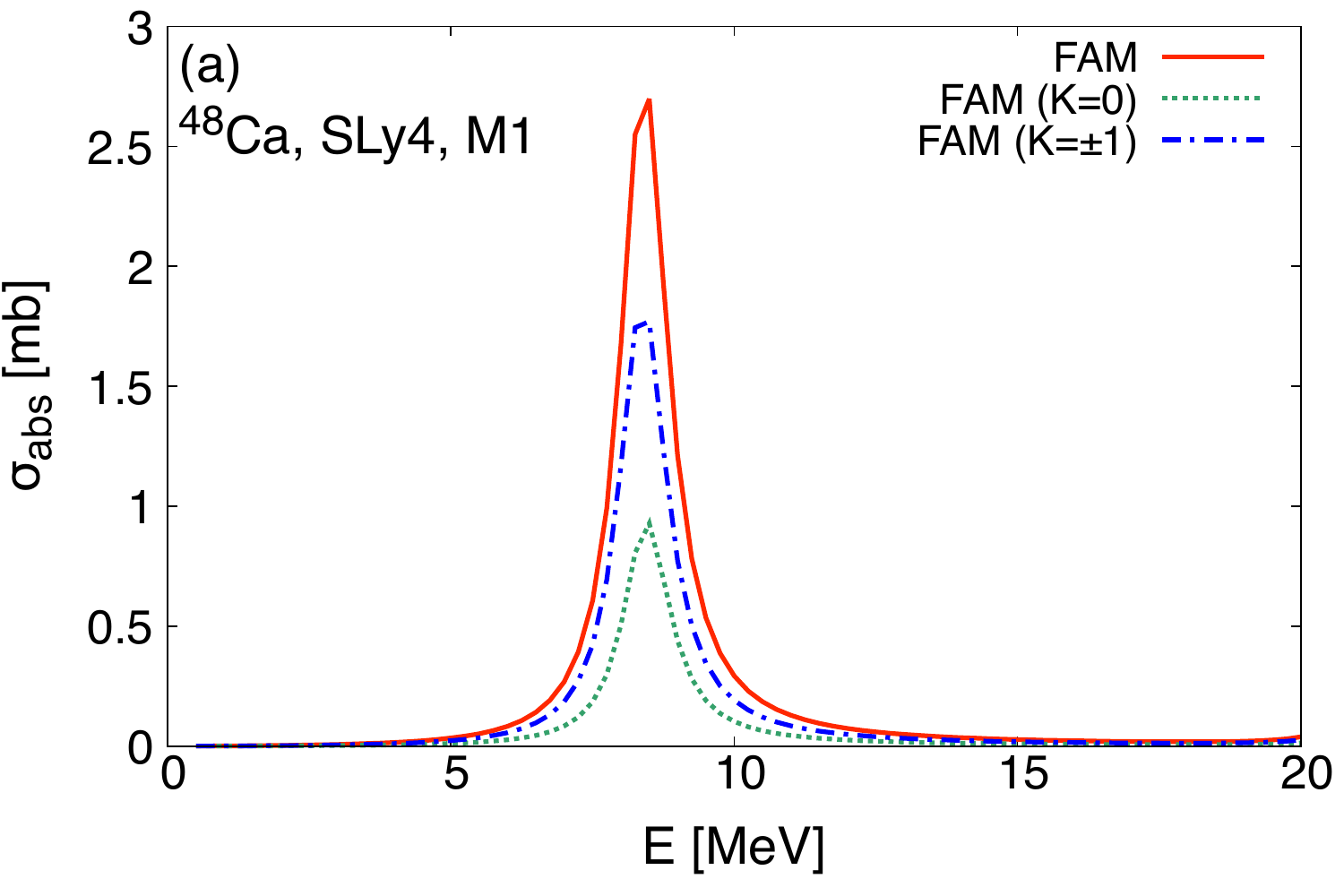}\\
\includegraphics[width=1\linewidth]{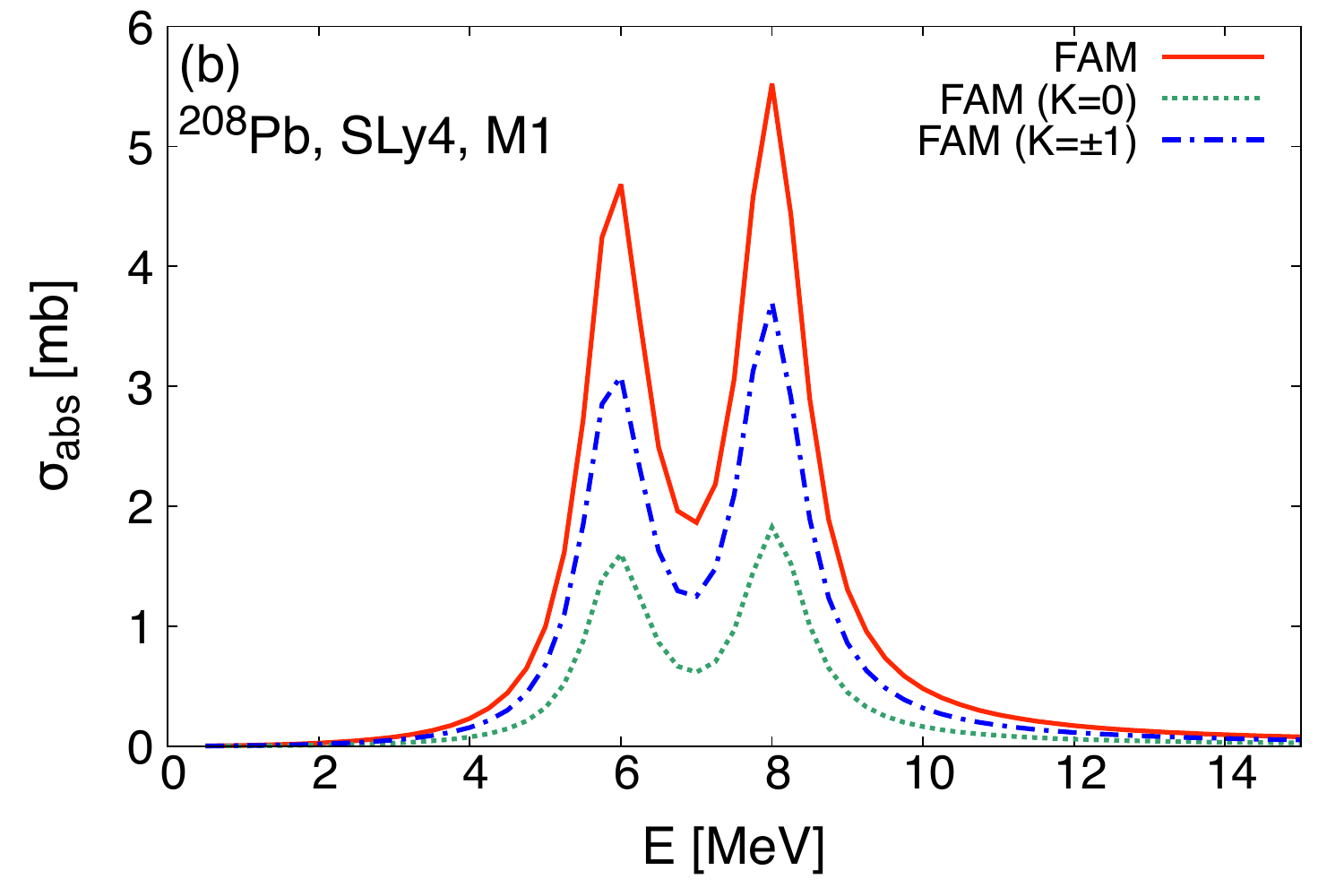}
\caption{
The photoabsorption cross sections of M1 transitions for double-magic nuclei, (a) $^{48}\mathrm{Ca}$ and (b) $^{208}\mathrm{Pb}$. The solid lines show the results of Eq.~(\ref{eq:cross section M1}). The partial contributions from $K=0$ and $K=\pm1$ to the total strength are shown by the dotted and dash-dotted lines.
}
\label{fig:M1 double magic}
\end{figure}

For the M$1$ transition, we use the magnetic dipole operator in Eq. (\ref{eq:magnetic dipole operator}) to perform almost the same calculation as done in Sec.\ref{sec:E1 transition}. The value of the Lorentzian width is fixed to $\gamma=1$ MeV. Here, we focus on the M$1$ transitions for double-magic nuclei.

\begin{figure}[htbp]
\includegraphics[width=1\linewidth]{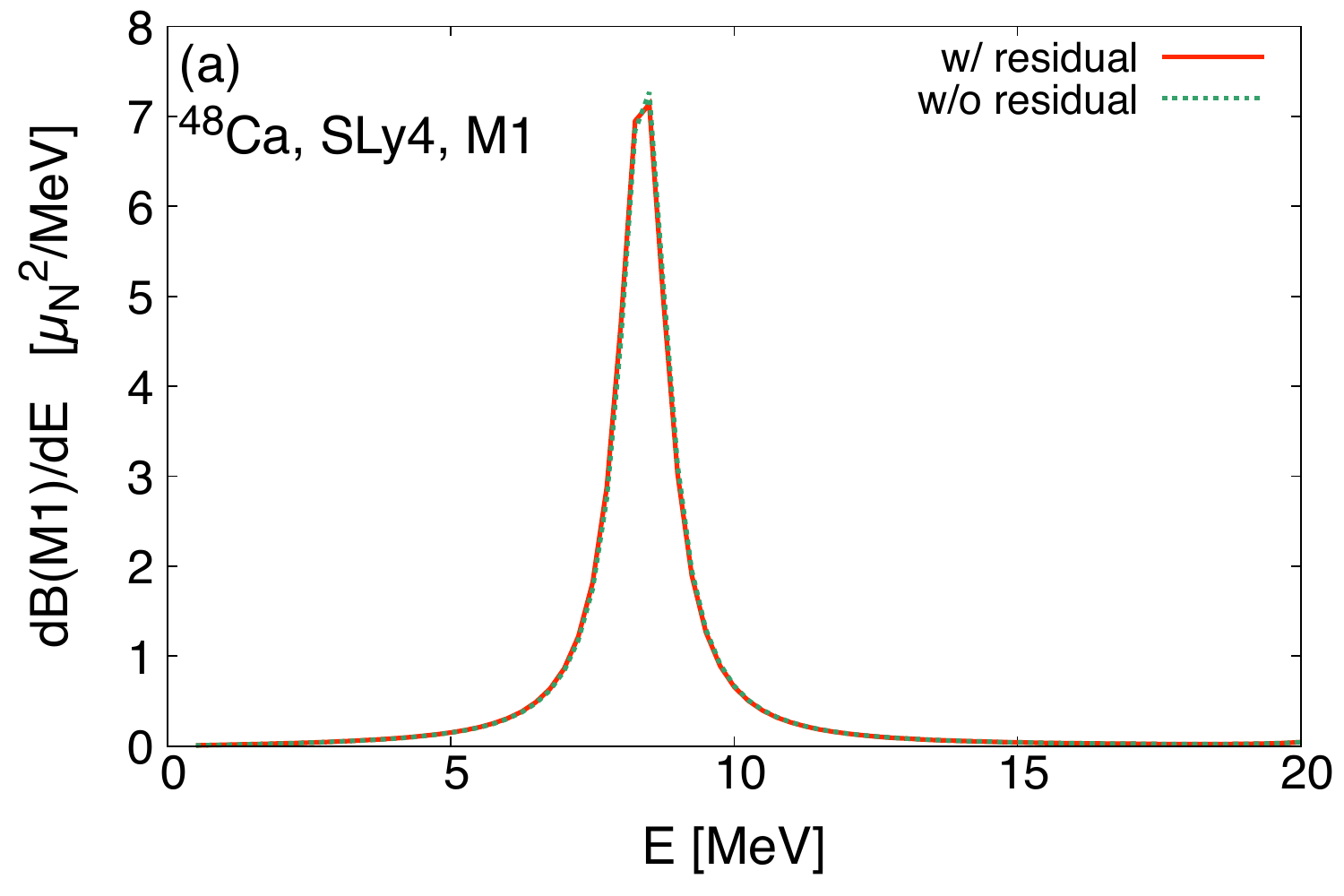}\\
\includegraphics[width=1\linewidth]{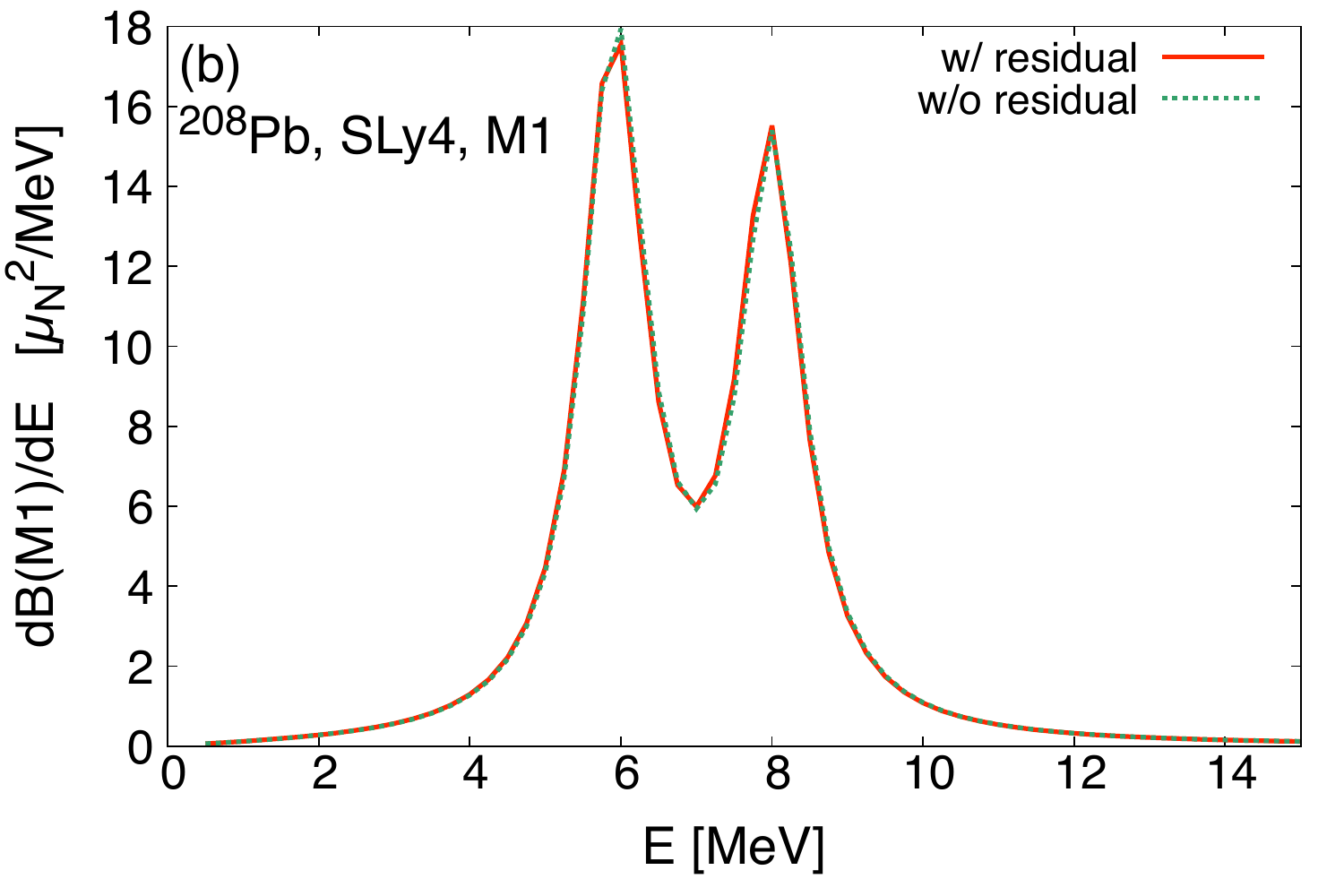}
\caption{
The transition strength distributions in Eq.~(\ref{eq:transition strength distribution M1}) for (a) $^{48}\mathrm{Ca}$ and (b) $^{208}\mathrm{Pb}$. The solid (dotted) line shows the result with (without) the residual interactions in Eqs.~(\ref{eq:partial derivative time-even Hamiltonian}) and (\ref{eq:partial derivative time-odd Hamiltonian}).
}
\label{fig:M1 double magic strength}
\end{figure}

Figure~\ref{fig:M1 double magic} shows the cross sections of M$1$ transitions for $^{48}\mathrm{Ca}$ and $^{208}\mathrm{Pb}$. The transition strengths for $K=0,\pm1$ are almost identical. In the case of these double-magic nuclei, the spin part in Eq.~(\ref{eq:magnetic dipole operator}) dominantly contributes to the transition strength of the M$1$ transitions. 

 The dominant peak around $8.5$ MeV in Fig.~\ref{fig:M1 double magic}(a) is originated from a spin-flip $1f_{7/2}\to1f_{5/2}$ transition of neutrons \cite{Kruzic:2020oqf}. As shown in Fig.~\ref{fig:M1 double magic}(b), there are double peaks in the M$1$ transition of $^{208}\mathrm{Pb}$. These two peaks are not by the different $K$ values but due to contributions from neutron spin-flip transitions and proton spin-flip transitions. The peak around $8.0$ MeV ($6.0$ MeV) in Fig.~\ref{fig:M1 double magic}(b) reflects the contribution from a spin-flip $1i_{13/2}\to1i_{11/2}$ ($1h_{11/2}\to1h_{9/2}$) transition of neutrons (protons), respectively. In our FAM calculations, the M$1$ transition strength distribution is given by
\begin{equation}
\label{eq:transition strength distribution M1}
    \frac{\mathrm{d}B(\mathrm{M}1)}{\mathrm{d}E}=\sum_{K=0,\pm1}\frac{\mathrm{d}B(E;M_{K})}{\mathrm{d}E},
\end{equation}
where the right hand side of the above equation is calculated from Eqs.~(\ref{eq:strength function}) and (\ref{eq:magnetic dipole operator}). The results of such distributions for $^{48}\mathrm{Ca}$ and $^{208}\mathrm{Pb}$ are shown in the solid lines of Fig.~\ref{fig:M1 double magic strength}. The strength distribution of $^{48}\mathrm{Ca}$ is well consistent with that of Ref.~\cite{Kruzic:2020oqf}. On the other hand, experiments of inelastic scatterings such as $^{48}\mathrm{Ca}(e,e^{\prime})$\cite{Steffen:1980b} and $^{48}\mathrm{Ca}(p,p^{\prime})$ \cite{Birkhan:2013vsa} found a dominant peak of the M$1$ excitation at $10.23$ MeV that is higher than the peak energy of $8.5$ MeV in Fig.~\ref{fig:M1 double magic}(a). As shown in Fig.~\ref{fig:M1 double magic strength}, the results of M$1$ transitions with the residual interactions (solid lines) are identical to those without residual interactions (dot lines), which indicates residual interactions in Eqs.~(\ref{eq:partial derivative time-even Hamiltonian}) and (\ref{eq:partial derivative time-odd Hamiltonian}) do not contribute to the M$1$ transitions. Here, we employ SLy$4$ as the Skyrme force that ignores the contribution from the spin terms with $\Tilde{b}_{i},\Tilde{b}_{i}^{\prime}$ \cite{Vesely:2009eb,Nesterenko:2010ra}. These spin terms may increase the repulsive effects of residual interactions and upshift the peak energy of the M$1$ transitions. 

Figure~\ref{fig:M1 double magic strength}(b) shows the double-peak feature of M$1$ for $^{208}\mathrm{Pb}$, which is also confirmed by other theoretical works in the past, nevertheless the experimental data show only one peak near $7.3$ MeV \cite{Birkhan:2013vsa}. Such discrepancy could be attributed to theoretical uncertainty in Skyrme parameters, because some other works, which adopt different Skyrme parameterizations (e.g. SkO and SkO’), reported single peak shapes \cite{Vesely:2009eb,Nesterenko:2010ra}. The result of the M$1$ transition is more sensitive to the Skyrme parameterization than that of the E$1$ transition.

The total transition strength $\sum B(\rm{M}1)$ is commonly used to compare calculated results with experimental data. In our FAM-RPA calculation, the total strength can be estimated from the integration of Eq.~(\ref{eq:transition strength distribution M1}) over the energy $E$,
\begin{equation}
\label{eq:total transition strength M1}
\begin{split}
    \sum B(\rm{M}1)&
    =\int\mathrm{d}E\frac{\mathrm{d}B(\mathrm{M}1)}{\mathrm{d}E},
\end{split}
\end{equation}
and they are 11.7~$\mu_{N}^{2}(32.5\mu_{N}^{2})$ for $^{48}\mathrm{Ca}(^{208}\mathrm{Pb})$, which are larger than the experimental data, e.g., 3.85 -- 4.63~$\mu_{N}^{2}$(20.5~$\mu_{N}^{2})$ for $^{48}\mathrm{Ca}(^{208}\mathrm{Pb})$~\cite{Birkhan:2013vsa}. 
It is also reported that the calculated M1 transition in the past tends to give larger $\sum B({\rm M1})$ (e.g. \cite{harakeh2001giant} and references therein), and they suggested to introduce the quench of the spin $g$ factors. The spin-flip part is dominant in the M$1$ transitions of double-magic nuclei so that the total strength in Eq.~(\ref{eq:total transition strength M1}) is almost proportional to the square of the quenching factor. By introducing a typical value of the quenching factor, $g^{(i)}_{s,\mathrm{eff}}/g^{(i)}_{s}=$~0.6 -- 0.7~\cite{Kruzic:2020oqf}, the total strength in our FAM calculations is reduced by 0.36 -- 0.49 times and more consistent with the experimental data \cite{Birkhan:2013vsa}. 

When the nucleus is strongly deformed like nuclei in the rare-earth or actinide region, the so-called scissors mode may appear near 3~MeV~\cite{Heyde:2010ng}, which comes from the orbital part in Eq.~(\ref{eq:magnetic dipole operator}). They are very sensitive to the low-excitation configurations near the Fermi surface, and unfortunately our current implementation of non-iterative FAM-RPA brings a large uncertainty in calculating the scissors mode. This issue could be resolved by our next modeling of non-iterative FAM-QRPA.


\section{Conclusion}\label{sec:Conclusion}


We develop a non-iterative FAM method, where the forward and backward amplitudes are derived from the explicit linearization of the residual interaction without an iterative procedure used in the other conventional FAM-RPA calculations. We carried out the HF+BCS calculation to prepare the ground state of the static HF Hamiltonian, then the occupied and unoccupied states are fed into the calculation of the E$1$ transitions for both the spherical and deformed nuclei. Our calculated result indicates that the E$1$ transition is not so sensitive to the detail of the single-particle states near the Fermi surface.

We also applied our FAM-RPA calculations to the M$1$ transitions for double-magic nuclei and demonstrated that the spin-flip of neutrons and protons is the main contribution to the M$1$ transition. FAM-RPA tends to overestimate the M$1$ transition strength, which might be resolved when we introduce quenching of the spin $g$ factor. We also discussed the sensitivity of the spin term in the Skyrme force to the M$1$ transition, which is currently neglected in our calculations. This could shift the M$1$ resonance location higher, making our calculations more consistent with the experimental data.

\begin{acknowledgments}
This work was partially support by the Office of Defense Nuclear Nonproliferation Research \& Development (DNN R\&D), National Nuclear Security Administration,
U.S. Department of Energy. This work was carried out under the
auspices of the National Nuclear Security Administration of the
U.S. Department of Energy at Los Alamos National Laboratory under
Contract No.~89233218CNA000001.
\end{acknowledgments}

\appendix

\section{Integrands in the residual interaction}
\label{sec:detail of integrands}

We show the detailed description of integrands inside Eqs. (\ref{eq:partial derivative time-even Hamiltonian}) and (\ref{eq:partial derivative time-odd Hamiltonian}). To prepare the HF single-particle state for the FAM-RPA calculation, we carry out the HF+BCS calculation as done in Ref.~\cite{Bonneau:2007dc}. The HF single-particle state on the right-hand sides of Eqs.~(\ref{eq:single-particle states TDHF}) and (\ref{eq:single-particle states TDHF complex conjugate}) is described in the cylindrical coordinate space $\vec{r}=(r,\varphi,z)$ \cite{Vautherin:1973zz},
\begin{equation}
\label{eq:single-particle states HF}
\begin{split}
\phi_{\mu}^{q}
=\chi_{q}(\tau)\left\{
\phi^{+}_{\mu}(r,z)\chi_{1/2}(\sigma)\frac{e^{i\Lambda^{-}_{\mu}\varphi}}{\sqrt{2\pi}}\right.\left.+\phi^{-}_{\mu}(r,z)\chi_{-1/2}(\sigma)\frac{e^{i\Lambda^{+}_{\mu}\varphi}}{\sqrt{2\pi}}\right\},
\end{split}
\end{equation}
\begin{equation}
    \Lambda^{\pm}_{\mu}=\Omega_{\mu}\pm\frac{1}{2},
\end{equation}
where $\sigma(\tau)$ represents the spin (isospin) of the nucleon. $\chi_{q}(\tau)$ is the eigenstate of the isospin operator $\tau_{z}$ and the index $q$ labels neutrons or protons. The $\chi_{1/2}(\sigma)$ and $\chi_{-1/2}(\sigma)$ represent the up and down states of the nucleon spin, respectively. The wave function $\phi^{\pm}_{\mu}(r,z)$ is expanded by the cylindrical harmonic oscillator basis. The $\Omega_{\mu}$ represents the projection of the total angular momentum on $z$-axis. The integrands such as $\phi^{q*}_{m}\phi_{i}^{q}$, $\vec{\nabla}\phi^{q*}_{m}\vec{\nabla}\phi_{i}^{q}$, $\phi^{q*}_{m}\nabla^{2}\phi_{i}^{q}$, and $(-i)\vec{\nabla}\phi^{q*}_{m}(\vec{\nabla}\times\vec{\sigma})\phi_{i}^{q}$ in Eq.~(\ref{eq:partial derivative time-even Hamiltonian}) are described in the cylindrical coordinate $(r,\varphi,z)$. The spin and isospin states inside these integrands are eliminated owing to the inner products, $\chi_{\Sigma^{\prime}}^{\dagger}(\sigma)\chi_{\Sigma}(\sigma)=\delta_{\Sigma\Sigma^{\prime}}\ (\Sigma,\Sigma^{\prime}=\pm1/2)$ and $\chi^{\dagger}_{q}(\tau)\chi_{q}(\tau)=1$. Similar to the detailed calculation of local densities and currents in Ref. \cite{Vautherin:1973zz}, the integrand functions in the residual interaction of the time-even Hamiltonian are written as
\begin{widetext}
\begin{equation}
\label{eq:pseudo density}
\begin{split}
\phi^{q*}_{m}\phi_{i}^{q}=\frac{e^{i(\Omega_{i}-\Omega_{m})\varphi}}{2\pi}\left(
\phi^{+}_{m}\phi^{+}_{i}+\phi^{-}_{m}\phi^{-}_{i}
\right),
\end{split}
\end{equation}

\begin{align}
\label{eq:pseudo kinetic density}
\vec{\nabla}\phi^{q*}_{m}\cdot\vec{\nabla}\phi_{i}^{q}=\frac{e^{i(\Omega_{i}-\Omega_{m})\varphi}}{2\pi}&\left\{
\nabla_{r}\phi^{+}_{m}\nabla_{r}\phi^{+}_{i}+\nabla_{z}\phi^{+}_{m}\nabla_{z}\phi^{+}_{i}
+\frac{\Lambda_{m}^{-}\Lambda_{i}^{-}}{r^{2}}\phi^{+}_{m}\phi^{+}_{i}\right. \notag \\
&+\left.\nabla_{r}\phi^{-}_{m}\nabla_{r}\phi^{-}_{i}+\nabla_{z}\phi^{-}_{m}\nabla_{z}\phi^{-}_{i}
+\frac{\Lambda_{m}^{+}\Lambda_{i}^{+}}{r^{2}}\phi^{-}_{m}\phi^{-}_{i}
\right\},
\end{align}

\begin{equation}
\label{eq:Laplacian funtion}
\begin{split}
\phi^{q*}_{m}\nabla^{2}\phi_{i}^{q}=\frac{e^{i(\Omega_{i}-\Omega_{m})\varphi}}{2\pi}&\left\{
\phi^{+}_{m}\left[
\frac{1}{r}\nabla_{r}(r\nabla_{r})
+\nabla_{z}^{2}-\frac{(\Lambda_{i}^{-})^{2}}{r^{2}}
\right]\phi_{i}^{+}
+\phi^{-}_{m}\left[\frac{1}{r}\nabla_{r}(r\nabla_{r})
+\nabla_{z}^{2}-\frac{(\Lambda_{i}^{+})^{2}}{r^{2}}
\right]\phi_{i}^{-}
\right\},
\end{split}
\end{equation}

\begin{align}
\label{eq:pseudo spin orbit density}
(-i)\vec{\nabla}\phi^{q*}_{m}(\vec{\nabla}\times\vec{\sigma})\phi_{i}^{q}=\frac{e^{i(\Omega_{i}-\Omega_{m})\varphi}}{2\pi}&\left\{
\nabla_{r}\phi_{i}^{+}\nabla_{z}\phi_{m}^{-}-\nabla_{r}\phi_{i}^{-}\nabla_{z}\phi_{m}^{+}+\nabla_{r}\phi_{m}^{+}\nabla_{z}\phi_{i}^{-}-\nabla_{r}\phi_{m}^{-}\nabla_{z}\phi_{i}^{+}\right. \notag\\
&+\left.\frac{\Lambda_{m}^{-}}{r}\phi_{m}^{+}(\nabla_{r}\phi_{i}^{+}-\nabla_{z}\phi_{i}^{-})+\frac{\Lambda_{i}^{-}}{r}\phi_{i}^{+}(\nabla_{r}\phi_{m}^{+}-\nabla_{z}\phi_{m}^{-})\right. \notag\\
&-\left.\frac{\Lambda_{m}^{+}}{r}\phi_{m}^{-}(\nabla_{r}\phi_{i}^{-}+\nabla_{z}\phi_{i}^{+})-\frac{\Lambda_{i}^{+}}{r}\phi_{i}^{-}(\nabla_{r}\phi_{m}^{-}+\nabla_{z}\phi_{m}^{+})
\right\},
\end{align}
\end{widetext}
where we use simple notations, $\nabla_{r(z)}\equiv\partial/\partial r(z)$ and  $\phi_{\mu}^{\pm}\equiv\phi^{\pm}_{\mu}(r,z)$. The operator $(\vec{\nabla}\times\vec{\sigma})$ in Eq.~(\ref{eq:pseudo spin orbit density}) is described in the cylindrical coordinate. $\vec{\sigma}$ operates the spin eigenstates as in $\sigma_{\pm}\chi_{1/2(-1/2)}=0$, $\sigma_{\pm}\chi_{-1/2(1/2)}=2\chi_{1/2(-1/2)}$, and $\sigma_{z}\chi_{1/2(-1/2)}=\pm\chi_{1/2(-1/2)}$. The $\phi_{\mu}^{\pm}$ is a real number function in the $(r,z)$ space. The detailed description of $\nabla_{r(z)}\phi_{\mu}^{\pm}$ is shown in Ref. \cite{Vautherin:1973zz}. All of the $\varphi$ dependence is summarized into the factor $e^{i(\Omega_{i}-\Omega_{m})\varphi}$ in Eqs.~(\ref{eq:pseudo density})-(\ref{eq:pseudo spin orbit density}). The spatial integrals in Eq.~(\ref{eq:partial derivative time-even Hamiltonian}) are carried out in the cylindrical coordinate $(r,\varphi,z)$. The integration towards the $\varphi$ direction induces a condition, $ \Omega_{i}-\Omega_{m}=\Omega_{j}-\Omega_{n}$ for the finite value of Eq.~(\ref{eq:partial derivative time-even Hamiltonian}). Such a condition of angular momentum restricts the size of the RPA matrix $A$. In the case of the RPA matrix $B$, the condition becomes $ \Omega_{i}-\Omega_{m}=-(\Omega_{j}-\Omega_{n})$, where the negative sign on the right-hand side comes from the relation in Eq.~(\ref{eq:RPA matrix B relation}). The integrands in Eq.~(\ref{eq:partial derivative time-even Hamiltonian}) can be expanded by a linear combination of Hermite (associated Laguerre) polynomials in the $z(r)$ direction by following the formalism in Ref.~\cite{Vautherin:1973zz}. The spatial integrations over $z$ and $r$ directions in Eq.~(\ref{eq:partial derivative time-even Hamiltonian}) can be carried out by the Gaussian quadratures. In the case of the time-odd Hamiltonian, the residual interaction in Eq.~(\ref{eq:partial derivative time-odd Hamiltonian}) includes two types of vectors, $\vec{\nabla}\times(\phi^{q*}_{m}\vec{\sigma}\phi_{i}^{q})$ and $\frac{1}{2i}(\phi^{q*}_{m}\vec{\nabla}\phi_{i}^{q}-\phi^{q}_{i}\vec{\nabla}\phi_{m}^{q^{*}})$. In the cylindrical coordinate, the components of these vectors are written as
\begin{widetext}
\begin{align}
\left\{\vec{\nabla}\times(\phi^{q*}_{m}\vec{\sigma}\phi_{i}^{q})
\right\}_{r}&=\frac{e^{i(\Omega_{i}-\Omega_{m})\varphi}}{2\pi}i\left\{
\frac{\Omega_{i}-\Omega_{m}}{r}(\phi_{m}^{+}\phi_{i}^{+}-\phi_{m}^{-}\phi_{i}^{-})
-\nabla_{z}(-\phi_{m}^{+}\phi_{i}^{-}+\phi_{m}^{-}\phi_{i}^{+})
\right\},\notag\\
\left\{\vec{\nabla}\times(\phi^{q*}_{m}\vec{\sigma}\phi_{i}^{q})
\right\}_{\varphi}&=\frac{e^{i(\Omega_{i}-\Omega_{m})\varphi}}{2\pi}\left\{
\nabla_{z}(\phi_{m}^{+}\phi_{i}^{-}+\phi_{m}^{-}\phi_{i}^{+})-\nabla_{r}(\phi_{m}^{+}\phi_{i}^{+}-\phi_{m}^{-}\phi_{i}^{-})
\right\},\notag\\
\left\{\vec{\nabla}\times(\phi^{q*}_{m}\vec{\sigma}\phi_{i}^{q})
\right\}_{z}&=\frac{e^{i(\Omega_{i}-\Omega_{m})\varphi}}{2\pi}i\left\{
-\frac{\Omega_{i}-\Omega_{m}}{r}(\phi_{m}^{+}\phi_{i}^{-}+\phi_{m}^{-}\phi_{i}^{+})+(\nabla_{r}+\frac{1}{r})(-\phi_{m}^{+}\phi_{i}^{-}+\phi_{m}^{-}\phi_{i}^{+})
\right\},
\end{align}

\begin{align}
\frac{1}{2i}(\phi^{q*}_{m}\vec{\nabla}\phi_{i}^{q}-\phi^{q}_{i}\vec{\nabla}\phi_{m}^{q^{*}})_{r}&=\frac{e^{i(\Omega_{i}-\Omega_{m})\varphi}}{2\pi}\frac{1}{2i}\left(
\phi_{m}^{+}\nabla_{r}\phi_{i}^{+}-\phi_{i}^{+}\nabla_{r}\phi_{m}^{+}+\phi_{m}^{-}\nabla_{r}\phi_{i}^{-}-\phi_{i}^{-}\nabla_{r}\phi_{m}^{-}
\right),\notag \\
\frac{1}{2i}(\phi^{q*}_{m}\vec{\nabla}\phi_{i}^{q}-\phi^{q}_{i}\vec{\nabla}\phi_{m}^{q^{*}})_{\varphi}&=\frac{e^{i(\Omega_{i}-\Omega_{m})\varphi}}{2\pi}\frac{1}{2}\left(
\frac{\Omega_{m}+\Omega_{i}-1}{r}\phi_{m}^{+}\phi_{i}^{+}
+\frac{\Omega_{m}+\Omega_{i}+1}{r}\phi_{m}^{-}\phi_{i}^{-}
\right),\notag\\
\frac{1}{2i}(\phi^{q*}_{m}\vec{\nabla}\phi_{i}^{q}-\phi^{q}_{i}\vec{\nabla}\phi_{m}^{q^{*}})_{z}&=\frac{e^{i(\Omega_{i}-\Omega_{m})\varphi}}{2\pi}\frac{1}{2i}\left(
\phi_{m}^{+}\nabla_{z}\phi_{i}^{+}-\phi_{i}^{+}\nabla_{z}\phi_{m}^{+}+\phi_{m}^{-}\nabla_{z}\phi_{i}^{-}-\phi_{i}^{-}\nabla_{z}\phi_{m}^{-}
\right),
\end{align}
\end{widetext}
where the dependence of $\varphi$ is included in the factor $e^{i(\Omega_{i}-\Omega_{m})\varphi}$. As in the case of the time-even Hamiltonian, $\Omega_{i}-\Omega_{m}=\Omega_{j}-\Omega_{n}$ is necessary for the finite value of the residual interaction in Eq.~(\ref{eq:partial derivative time-odd Hamiltonian}).

\section{Calculation of the Coulomb potential}
\label{sec:detail of Coulomb term}

The contribution from the direct term of the Coulomb potential in Eq.~(\ref{eq:partial derivative time-even Hamiltonian}) is written as 
\begin{align}
\label{eq:direct Coulomb}
(\Tilde{V}_{C})_{nj}&=\frac{e^{2}}{2}\int\mathrm{d}^{3}r^{\prime}\ \frac{\phi_{j}^{p*}\phi_{n}^{p}}{|\vec{r}-\vec{r^{\prime}}|} \notag\\
&=\frac{e^{2}}{4}\int\mathrm{d}^{3}r^{\prime}\ |\vec{r}-\vec{r^{\prime}}|\nabla^{\prime2}(\phi_{j}^{p*}\phi_{n}^{p}),
\end{align}
where the relation $\nabla^{\prime2}|\vec{r}-\vec{r^{\prime}}|=2/|\vec{r}-\vec{r^{\prime}}|$ is used. Then, we carry out the integration by parts twice. Eq.~(\ref{eq:direct Coulomb}) is a function in the cylindrical coordinate $(r,\varphi,z)$. The integration can be described by following the procedure to calculate the Coulomb potential in Ref.~\cite{Vautherin:1973zz},
\begin{equation}
\begin{split}
(\Tilde{V}_{C})_{nj}=&
\frac{e^{i(\Omega_{n}-\Omega_{j})\varphi}}{2\pi}e^{2}\int_{0}^{\infty}\mathrm{d}r^{\prime}
\int_{-\infty}^{\infty}\mathrm{d}z^{\prime}\\
&\times\left\{\sqrt{
(r+r^{\prime})^{2}+(z-z^{\prime})^{2}
}J(r^{\prime},z^{\prime}) I(\Omega_{n}-\Omega_{j},x)
\right\},
\end{split}
\end{equation}
\begin{equation}
\label{eq:J function}
    J(r^{\prime},z^{\prime})=(2\pi)e^{-i(\Omega_{n}-\Omega_{j})\varphi^{\prime}}\nabla^{\prime2}(\phi_{j}^{p*}\phi_{n}^{p}),
\end{equation}
\begin{equation}
\label{eq:I function}
    I(x,\Omega)=(-1)^{\Omega}\int_{0}^{\pi/2}\mathrm{d}\varphi^{\prime}\sqrt{1-x^{2}\sin^{2}\varphi^{\prime}}\cos(2\Omega\varphi^{\prime}),
\end{equation}
\begin{equation}
x^{2}=\frac{4rr^{\prime}}{\sqrt{
(r+r^{\prime})^{2}+(z-z^{\prime})^{2}
}},
\end{equation}
where the factor $e^{-i(\Omega_{n}-\Omega_{j})\varphi^{\prime}}$ in Eq.~(\ref{eq:J function}) cancels the $\varphi^{\prime}$ dependence in $\nabla^{\prime2}(\phi_{j}^{p*}\phi_{n}^{p})$ that is calculated from Eqs.~(\ref{eq:pseudo kinetic density}) and (\ref{eq:Laplacian funtion}). The difference in angular momentum is restricted to $|\Omega_{n}-\Omega_{j}|=0,1$ when the external fields are the operators of E$1$ and M$1$ transitions. At $\Omega=0$, Eq.~(\ref{eq:I function}) corresponds to the well-known complete elliptic integral of the second kind. After the Taylor expansion and term by term integration, Eq.~(\ref{eq:I function}) at $\Omega=0,\pm1$ can be expressed by the infinite power series,
\begin{align}
I(x,\Omega=0)&=\frac{\pi}{2}\sum_{n=0}^{\infty}\left\{
\frac{(2n-1)!!}{(2n)!!}
\right\}^{2}\frac{x^{2n}}{1-2n},\\
I(x,\Omega=\pm1)&=\frac{\pi}{2}\sum_{n=0}^{\infty}\left\{
\frac{(2n-1)!!}{(2n)!!}
\right\}^{2}
\frac{n}{(n+1)(1-2n)}x^{2n},
\end{align}
where $(-1)!!\equiv 1$. This expansions for $x^2 < 1$ converge sufficiently when $n=10$ or higher.

\section{Coefficients of the external fields}
\label{sec:detail of the coefficient of external fields}
In the case of E$1$ transition, the coefficient $f_{mi}^{q}$ in Eq.~(\ref{eq:external field f}) is calculated by
\begin{align}
f_{mi}^{q}&=\int\mathrm{d}^{3}r\ \phi_{m}^{q*}\sum_{K=0,\pm1}D_{K}\phi_{i}^{q} \notag\\
&=\sqrt{\frac{3}{4\pi}}\int_{0}^{\infty}\mathrm{d}r\ r\int_{-\infty}^{\infty}\mathrm{d}z\ g^{E1}(\Omega_{mi},q),
\end{align}

\begin{equation}
\label{eq:Omega condition E1}
    g^{E1}(\Omega_{mi},q)=\left\{
\begin{array}{cc}
ze_{\mathrm{eff}}^{q}(\phi^{+}_{m}\phi^{+}_{i}+\phi^{-}_{m}\phi^{-}_{i})&  (\Omega_{mi}=0)\\
\mp \frac{r}{\sqrt{2}}e_{\mathrm{eff}}^{q}(\phi^{+}_{m}\phi^{+}_{i}+\phi^{-}_{m}\phi^{-}_{i})& (\Omega_{mi}=\pm1) \\
0& (\mathrm{otherwise})
\end{array}
\right.,
\end{equation}
where $\Omega_{mi}\equiv\Omega_{m}-\Omega_{i}$ and $e_{\mathrm{eff}}^{q}\equiv-eZ/A(eN/A)$ for $q=n(p)$. The condition of $\Omega_{mi}$ comes from the integration over $\varphi$, which restricts the size of the configuration space. We can derive the coefficient $f_{im}^{q}$ in Eq.~(\ref{eq:external field f}) in the same way. In the case of M$1$ transition, the operator $\vec{\sigma}$ can flip the spin of nucleons. The coefficient $f_{mi}^{q}$ is given by
\begin{align}
f_{mi}^{q}&=\int\mathrm{d}^{3}r\ \phi_{m}^{q*}\sum_{K=0,\pm1}M_{K}\phi_{i}^{q}\\
&=\sqrt{\frac{3}{4\pi}}\int_{0}^{\infty}\mathrm{d}r\ r\int_{-\infty}^{\infty}\mathrm{d}z\ g^{M1}(\Omega_{mi},q),
\end{align}
where the $g^{M1}(\Omega_{mi},q)$ is finite only for $\Omega_{mi}=0,\pm1$ as in Eq.~(\ref{eq:Omega condition E1}). The detailed description of the function $g^{M1}(\Omega_{mi},q)$ is written as 
\begin{align}
g^{M1}(0,q)&=\mu_{N}g_{l}^{q}(\Lambda^{-}_{i}\phi_{m}^{+}\phi_{i}^{+}+\Lambda^{+}_{i}\phi_{m}^{-}\phi_{i}^{-}) \notag\\
&+\frac{\mu_{N}g_{s}^{q}}{2}(\phi_{m}^{+}\phi_{i}^{+}-\phi_{m}^{-}\phi_{i}^{-}),
\end{align}
\begin{align}
g^{M1}(1,q&)=\frac{\mu_{N}g_{l}^{q}}{\sqrt{2}}\phi_{m}^{+}(-z\nabla_{r}+r\nabla_{z})\phi_{i}^{+} \notag\\
&+\frac{\mu_{N}g_{l}^{q}}{\sqrt{2}}\phi_{m}^{-}(-z\nabla_{r}+r\nabla_{z})\phi_{i}^{-} \notag\\
&+\frac{\mu_{N}g_{l}^{q}}{\sqrt{2}}\frac{z}{r}(\Lambda_{i}^{-}\phi_{m}^{+}\phi_{i}^{+}+\Lambda_{i}^{+}\phi_{m}^{-}\phi_{i}^{-})-\frac{\mu_{N}g_{s}^{q}}{\sqrt{2}}\phi_{m}^{+}\phi_{i}^{-},
\end{align}
\begin{align}
g^{M1}(-1,q&)=\frac{\mu_{N}g_{l}^{q}}{\sqrt{2}}\phi_{m}^{+}(-z\nabla_{r}+r\nabla_{z})\phi_{i}^{+} \notag\\
&+\frac{\mu_{N}g_{l}^{q}}{\sqrt{2}}\phi_{m}^{-}(-z\nabla_{r}+r\nabla_{z})\phi_{i}^{-} \notag\\
&-\frac{\mu_{N}g_{l}^{q}}{\sqrt{2}}\frac{z}{r}(\Lambda_{i}^{-}\phi_{m}^{+}\phi_{i}^{+}+\Lambda_{i}^{+}\phi_{m}^{-}\phi_{i}^{-})+\frac{\mu_{N}g_{s}^{q}}{\sqrt{2}}\phi_{m}^{-}\phi_{i}^{+},
\end{align}
where $g_{s}^{q}=-3.826(5.586)$ and $g_{l}^{q}=0(1)$ for $q=n(p)$.


\bibliography{ref}


\end{document}